\documentclass[a4paper,11pt]{article}
\usepackage{amsmath,a4wide}
\usepackage{amssymb}
\usepackage{amscd}
\usepackage{epsfig}
\usepackage{graphics}
\usepackage[T1]{fontenc}
\usepackage{graphicx}
\usepackage{epsfig}
\usepackage{amsmath}
\usepackage{graphics}
\usepackage{graphicx}
\usepackage{epstopdf}
\usepackage{color}
\usepackage{booktabs}
\usepackage{subfigure}
\usepackage{ifpdf}
\usepackage{amssymb}
\usepackage{cite}
\usepackage{amssymb}
\usepackage{amsthm}
\usepackage{amsmath}
\usepackage[]{natbib}

\newtheorem{pro}{Proposition}[section]
\newtheorem{pr}{Problem}

\newtheorem{thm}{Theorem}[section]
\newtheorem{cor}{Corollary}[section]
\newtheorem{lem}{Lemma}[section]

\newtheorem{as}{Assumption}
\newtheorem{rem}{Remark}[section]

\def\t{\theta}

\def\z{\zeta}

\begin{document}
	
	\title{\LARGE {\bf Optimal Insurance with Limited Commitment in a Finite Horizon\footnote{
				Junkee Jeon gratefully acknowledges the support of the National Research Foundation of Korea (NRF) grant funded by the Korea government (Grant No. NRF-2017R1C1B1001811). Hyeng Kuen Koo gratefully acknowledges the support of the National Research Foundation of Korea (NRF) grant funded by the Korea government (MSIP) (Grant No. NRF-2016R1A2B4008240). Kyunghyun Park is supported by NRF Global Ph.D Fellowship (2016H1A2A1908911). }}}
	
	\author{
		Junkee Jeon \footnote{E-mail: {\tt junkeejeon@khu.ac.kr}\;Department of Applied Mathematics, Kyung Hee University, Korea.} 
		\and
		Hyeng Keun Koo\footnote{E-mail: {\tt hkoo@ajou.ak.kr}\;Department of Financial Engineering, Ajou University, Korea.}
		\and
		Kyunghyun Park\footnote{ E-mail: {\tt luminary9699@snu.ac.kr}\;Department of Mathematical Sciences, Seoul National University, Korea.}
	}
	
	\date{\today}
	
	\maketitle \pagestyle{plain} \pagenumbering{arabic}
	
	\abstract{We study a finite horizon optimal contracting problem with limited commitment. A risk-neutral principal enters into an insurance contract with a risk-averse agent who receives a stochastic income stream and is unable to make any commitment.  This problem involves an infinite number of constraints at  all times and at each state of the world. \citet{MJ} have developed a dual approach to the problem by considering a Lagrangian and derived a Hamilton-Jacobi-Bellman equation in an infinite horizon. We consider a similar Lagrangian in a finite horizon, but transform the dual problem into an infinite series of optimal stopping problems. For each optimal stopping problem we provide an analytic solution by providing an integral equation representation for the free boundary. We provide a verification theorem that the value function of the original principal's problem is the Legender-Fenchel transform of the integral of the value functions of the optimal stopping problems. We also provide numerical simulations results of  the optimal contracting strategies.}
	
	\vspace{1.0cm} {\em Keywords} : Optimal contract, Limited commitment, Principal-Agent problem, Optimal stopping problem, Variational inequality, Singular control problem \\

\newpage

\section{Introduction}

In this paper we investigate a contracting problem with limited commitment in continuous time. A risk-neutral principal enters into an insurance contract with a risk-averse agent who receives a stream of random income. The principal, typically a large institutional agent such as a government agency or a financial institution, is able to make firm commitment to keep the contract for reputational or for other reasons.  {However, the} agent{,} who may be an individual or a small country{,} is not able to make such   commitment. If both parties were able to make firm commitment, the contract would result in a classical outcome  {of} full insurance{, where} the principal absorbs all the risk in the agent's income and provides a constant stream of income to the agent. In the limited commitment case{,} the full insurance outcome is not attainable  {and} the principal provides only partial insurance to the agent. 

There has been intensive economic research on optimal contract {s} with limited commitment (see e.g.{,}  \citet{EG}, \citet{TW}, \citet{KL}).  Typically{,} the authors in the literature show that credit limits are used as a mechanism to enforce the contract and try to characterize the credit limits. They also show qualitative features of the contract. 
Recently, \citet{GZ} and \citet{MJ}  {proposed} formulating the problem in continuous time and obtaining a closed form characterization of the optimal contract.   The closed-form outcome exhibits the following feature of the optimal contract: the optimal contract starts with a payment much lower than in the full commitment case and ratchets up whenever the income process hits a new high level. They also show that the contract outcome can be enforced by providing unlimited insurance through futures contracts and  {optimal} imposing credit limits. 

We consider the continuous-time contracting problem with a finite contracting period. 
The infinite horizon models studied by \citet{GZ} and \citet{MJ} do not capture an important feature of  {real-world} contracts:  contracts mostly have finite maturity dates. This is the motivation of our paper.

We use the dual approach which allows us to use the Lagrangian method to solve the optimization problem.  {To} apply th {is} approach, we need to transform an infinite number of constraints into one constraint. We utilize a method proposed by \citet{HP} or \citet{MJ}. We construct the Lagrangian and define the dual problem by using the state variable {, i.e.,} which is the agent's income process{,} and a  costate variable{,}  {i.e.,} the cumulative Lagrange multiplier process  arising from the dynamic participation constraints. We solve the dual problem by transforming the problem into a series of optimal stopping problems. The optimal stopping problems are, in a formal sense,  equivalent to those of early exercise of American options. Then, we can apply well-developed techniques to the  {latter} contracting problem. In particular, we apply the integral representation of an American option value (see e.g. \citet{K}, \citet{J}, \citet{CJM}, \citet{Det}) and derive the optimal contract in analytic form.  {To} obtain a concrete solution, we apply the recursive integration method proposed by \citet{Huang} to solve numerically the integral equation. \\

\noindent {\it Contributions.}
Our contributions are as follows. First, we provide an analytic solution to the optimal contracting problem with limited commitment with a finite contracting period. Second, we  make a technical contribution, providing a connection between the contracting problem and the irreversible investment problem involving real options studied by \citet{DixitPindyck}. \citet{MJ} establish the duality theorem and provide a dynamic programming characterization of the dual problem. However, it is difficult to apply their method to the problem with a finite period, since one has to consider a Hamilton-Jacobi-Bellman(HJB) equation with a gradient constraint involving three variables {:} time, income, and the agent's continuation value. Moreover, it is not easy to find a solution to the HJB equation. We overcome th {is} difficulty by considering a transformed problem, which is similar in its formal structure to an irreversible incremental problem and equivalent to an infinite series of optimal stopping problems,  {so that it} essentially becomes a single optimal stopping problem in our model.\\

\noindent{\it Related literature.} In addition to the research mentioned above, there is extensive literature on the contracting problem with limited commitment. \citet{KL} and \citet{AJ2000} investigate asset pricing based on the model with limited commitment. 
\citet{Zhang} provides a solution to a long-term contracting problem in  discrete time using a stopping-time approach. \citet{BWY} shows that there is equivalence between the household problem and the contracting problem. \citet{AH2014}, \citet{AL}, \citet{AKL}, and \citet{BWY} study optimal  contracting between investors and an entrepreneur in a continuous-time model.
Our paper, however, is different from the papers in the literature in the following aspects. First, the contract maturity is finite in our continuous-time model, whereas most papers consider either discrete-time long-horizon or continuous-time infinite-horizon models. Second, most authors use dynamic programming{,} and \citet{MJ} use  {in particular} the dual approach, whereas we also use the dual approach  {as well as} transform {ing the original problem} into optimal stopping problems. The duality approach has been  used to study  continuous-time portfolio selection problems (see e.g., \citet{CoxH}, \citet{KLS}).  \citet{Jeon} also apply the dual approach and the transformation to study an  optimal consumption and portfolio selection problem in which an economic agent does not tolerate a decline in standard of living. \\

\noindent {\it Organization.} The rest of paper is organized as follows. Section \ref{sec:2} explains the model of the optimal contracting problem with limited commitment in a finite-horizon framework. In Section \ref{sec:3}, we state our optimization problem. By constructing the Lagrangian of the optimization problem, we define the dual problem. {In  {S}ection \ref{sec:B}, we analyse the variational inequality arising from the dual problem. Section \ref{sec:duality} establishes the duality theorem and provides the analytic representation of the optimal strategies.} In  {S}ection \ref{sec:4}, we provide numerical simulation results. In Section \ref{sec:5} we draw conclusions.
\section{Economic model}\label{sec:2}

We extend a contracting model with limited commitment in a continuous-time framework studied in \citet{GZ},  {and} \citet{MJ}. The crucial difference between the existing models and ours is that our model is set up in the finite horizon to study how the horizon affect {s} the contract.

The agent receives an income stream $Y_t$ which is an $\mathcal{F}_t$-adapted process. We consider the following geometric Brownian motion for the income process:
$$
dY_t= \mu Y_t dt+ \sigma Y_t dB_t,\;\;\;\;\;Y_0=y,
$$  
where $\mu>0, \sigma>0$ is constant and $\{B_t\}_{t=0}^{T}$ is a standard Brownian motion on the underlying probability space $(\Omega, \mathcal{F}, \mathbb{P})$ and $\{\mathcal{F}_t\}_{t=0}^{T}$ is the augmentation under $\mathbb{P}$ of the natural filtration generated by the standard Brownian motion $\{B_t\}_{t=0}^{T}$. The agent's income process $Y$ is publicly observable by both the principal and the agent.

The agent is risk-averse with subject discount rate $\rho>0$. The contract horizon is $[0,T]$. At time $0$, the principal offers  {a} contract $(C,T)$  to the agent. Then the instantaneous payment or consumption of the agent $C=\{C_t\}_{t=0}^{T}$ is a non-negative $\mathcal{F}_t$-adapted process satisfying 
$$
\mathbb{E}\left[\int_0^T e^{-rt}C_t dt\right] < \infty.
$$
Note that the contract is dependent on the  {entire} history of  {the} $Y_t$ process in our model.

We assume that the agent has no access to the financial market. The agent's utility at time $0$ is defined by 
$$
U_0^a(C) \equiv \mathbb{E}\left[\int_0^T e^{-\rho t} u(C_t) dt \right]{,}
$$  
where $u(\cdot)$ is a continuously differentiable, strictly concave, and strictly increasing function. Thus, the agent's continuation utility at date $t$ {is} given by 
$$
U_t^a(C) \equiv \mathbb{E}_t \left[ \int_t^{T} e^{-\rho(s-t)} u(C_s)ds \right]{,}
$$
where $\mathbb{E}_t[\cdot]=\mathbb{E}[\cdot \mid \mathcal{F}_t]$ is the conditional expectation at time $t$ on the filtration $\mathcal{F}_t$.

We assume that the principal can freely access the financial market with constant risk-free rate $r>0$ and can derive utility according to 
$$
U^p(y,C) \equiv \mathbb{E}\left[\int_0^Te^{-rt}(Y_t-C_t)dt\right].
$$  
Without loss of generality{,} we assume $0<r\le \rho$. This also means that the principal is more patient than or equally patient with the agent.

{To obtain} explicit solutions, we assume that the agent  {uses} the the constant relative risk aversion (CRRA) utility function as a representative utility function.
\begin{as}\label{as:1}
	\begin{eqnarray}\label{eq:CRRA}
	u(c)\equiv \frac{c^{1-\gamma}}{1-\gamma},~~~~~~\gamma>0,\gamma \neq 1{,}
	\end{eqnarray}
	where $\gamma$ is the agent's coefficient of relative risk aversion.
\end{as} 
The agent has a limited commitment. He/she can walk away from the contract and take an outside value at any time after signing the contract. The outside value (or autarky value) $U_d$ at time $t$ is given by
\begin{eqnarray*}
	\begin{split}
		U_d(t,Y_t) \equiv& \mathbb{E}_t \left[\int_t^T e^{-\rho(s-t)}u(Y_s)ds\right]
		=\dfrac{{(Y_t)}^{1-\gamma}}{1-\gamma}\dfrac{1-e^{-\hat{\rho}(T-t)}}{\hat{\rho}},
	\end{split}
\end{eqnarray*}
where $\hat{\rho}$ is defined by 
$$
\hat{\rho}\equiv \rho-(1-\gamma)\mu+\frac{1}{2}\gamma(1-\gamma)\sigma^2.
$$
We make the following assumption to ensure  a finite outside value for sufficiently large $T$.
\begin{as}
	$$\hat{\rho}>0.$$
\end{as}

To ensure that the agent does not walk away, we impose the following dynamic participation constraint:
\begin{eqnarray}\label{eq:EN2}
U_t^a(C) \geq U_d(t,Y_t),\;\;\;\;\forall t \in[0,T].
\end{eqnarray}\
We also impose the promise keeping constraint (or individual rationality constraint):
\begin{eqnarray}\label{eq:EN1}
w_0=U_0^a(C){,}
\end{eqnarray}
where $w_0$ is an initial promised value to the agent.
\begin{as}\label{as:promise_value}~
	We assume that the initial promised value $w$ satisfies the following inequality:
	$$
	w \ge U_d(0,y).
	$$
\end{as}
Lastly, the continuation utility of the principal at date $t$ is defined by 
$$
U_t^p({Y},C) \equiv \mathbb{E}_t\left[\int_t^Te^{-r(s-t)}(Y_s-C_s)ds\right].
$$
We call a consumption plan $\{C_s\}_{s=t}^{T}$ \textit{enforceable} at time $t\in[0,T]$ if the following conditions hold. 
\begin{itemize}
	\item[(i)](Integrability condition)
	\begin{eqnarray}\label{eq:int}
	\mathbb{E}_t\left[\int_t^T e^{-r(s-t)}C_s ds\right] < \infty.
	\end{eqnarray}
	\item[(ii)](Participation constraint)
	\begin{eqnarray}\label{eq:parti}
	U_s^a(C) \geq U_d(s,Y_s),~~~\forall s \in[t,T].
	\end{eqnarray}
\end{itemize}
Let $\Gamma_t(y,w)$ be the set of all \textit{enforceable} consumption plans at time $t$.

For some technical aspects, we make the following assumption:
\begin{as}\label{thm:as1}
	\begin{eqnarray}
	\begin{aligned}
	\hat{r}\equiv r-\mu >0,\;\;K\equiv r+\frac{\rho-r}{\gamma}~>0,~~{\textrm{and}}~~\mu>\frac{\sigma^2}{2}.
	\end{aligned}
	\end{eqnarray}
\end{as}
\section{ Optimization Problem}\label{sec:3}

\subsection{First-Best Allocation}
We first consider the first-best allocation contracting problem without participation constraint \eqref{eq:EN2} as the first-best benchmark. We use the agent's continuation value $w_t\equiv U_t^a(C)$ as a state variable.
\begin{pr}[First-Best Problem]\label{pro:FB} ~\\
	For given $w_t=w,\;{Y_t=y}$, the principal's problem is to maximize
	\begin{equation*}
	V^{FB}(t, y, w)= \sup_{\Gamma_t'(y,w)}U_t^p(y,C),
	\end{equation*}
	where $\Gamma'_t(y,w)$ is the set of all consumption plans $\{C_s\}_{s=t}^T$ satisfying
	the integrability condition \eqref{eq:int}.
\end{pr}
Note that the actual principal's problem for first-best case is to find $V^{FB}(0,y,w)$ at $t = 0$. However, the value function of Problem \ref{pro:FB} is well-defined for any $t > 0$ by the dynamic programming principle (\citet{Bellman})
since the principal fully commits to the contract until $T$.

For Lagrangian multiplier $\lambda^{*}>0$, let us consider the {following} Lagrangian for {the} first-best case: 
\begin{eqnarray*}
	\begin{split}
		{\bf L}_{FB}\equiv&\;\mathbb{E}_{t}\left[\int_{t}^{T}e^{-r(s-t)}(Y_s-C_s)ds\right]+\lambda^{*}\left(\mathbb{E}_{t}\left[\int_{t}^{T}e^{-\rho(s-t)}u(C_s)ds\right]-w\right).
	\end{split}
\end{eqnarray*}
Define the \textit{convex dual function} of $u$ as follows
\begin{eqnarray}\label{eq:DUAL}
\tilde{u}(z) \equiv \max_{c>0} \left\{zu(c)-c\right\},~~~\textrm{for}~~z>0.
\end{eqnarray}
In the case of the CRRA utility function, the dual function is derived as follows:   
\begin{eqnarray}\label{eq:CRRA_D}
\tilde{u}(z) =\frac{\gamma}{1-\gamma}z^{\frac{1}{\gamma}}.
\end{eqnarray}
Here, the first-best consumption is given by 
\begin{equation}\label{FB}
C_{s}^{FB}=(u')^{-1}\left({e^{(\rho-r)(s-t)}}/\lambda^{*}\right),\;\;\;\;s\in[t,T].
\end{equation}
From Lagrangian ${\bf L}_{FB}$, we can define the dual value function {$\tilde{V}^{FB}(t,\lambda^{*},y)$} as follows:
\begin{eqnarray*}
	\begin{split}
		\tilde{V}^{FB}(t,\lambda^{*},y)=&\;\mathbb{E}_{t}\left[\int_{t}^{T}e^{-r(s-t)}\tilde{u}\left(e^{-(\rho-r)(s-t)}\lambda^{*}\right)ds\right] +\mathbb{E}_{t}\left[\int_{t}^{T}e^{-r(s-t)}Y_s ds\right].
	\end{split}
\end{eqnarray*}
Then, we can deduce the following duality-relationship:
$$
V^{FB}(t,y,w)=\inf_{\lambda^{*}>0}\left(\tilde{V}^{FB}(t,\lambda^{*},y)-w\lambda^{*}\right).
$$
By a first-order condition, we can obtain 
$$
w= \dfrac{1-e^{-K(T-t)}}{K} \dfrac{1}{1-\gamma}(\lambda^{*})^{\frac{1}{\gamma}-1}.
$$
From \eqref{FB}, 
\begin{eqnarray}
\begin{split}\label{eq:FB_C}
C^{FB}_{s}=&e^{-\frac{(\rho-r)}{\gamma}(s-t)}\left(\dfrac{K(1-\gamma)w}{{1-e^{-K(T-t)}}}\right)^{\frac{1}{1-\gamma}},\qquad s\in[t,T].
\end{split}
\end{eqnarray}
For the first-best consumption $C^{FB}$, the following equality holds:
$$
U_0^a(C^{FB})=w.
$$
This means that the risk-neutral principal bears all uncertainty and fully insures the risk-averse agents.  

{Since $0<r\le \rho$,} we can easily confirm that the first-best consumption process $C^{FB}$ is non-increasing function over time.
\subsection{Limited Commitment}
We now write down the optimal contract problem with limited commitment between the principal and the agent.
\begin{pr}[Primal problem]\label{pr:main}~
	Given $w_t=w$ and {$Y_t=y$}, we consider the following maximization problem:
	$$
	V(t,y,w) \equiv \sup_{C \in \Gamma_t(y,w)}U_t^p(y,C).
	$$
\end{pr}
To obtain the solution of Problem \ref{pr:main}, we will construct a Lagrangian for Problem \ref{pr:main}. The key is to write the part of the Lagrangian corresponding to the participation constraint \eqref{eq:parti}, which should hold at every $s\in[t,T]$ and thus is comprised of infinite individual constraints. By utilizing the similar method proposed by \citet{HP}, \citet{MJ}  {wrote} an infinite number of constraints as an integral of the constraints. We write down the Lagrangian as follows:
\begin{eqnarray}\label{eq:LGR}
\begin{aligned}
{\bf L} \equiv & \mathbb{E}_t\left[ \int_t^T e^{-r(s-t)}  (Y_s-C_s)ds \right] + \lambda \left(\mathbb{E}_t\left[\int_t^T e^{-\rho(s-t)} u(C_s)ds\right]-w\right)\\
+&\mathbb{E}_t\left[\int_t^T e^{-r(s-t)}  \eta_s \left(\int_s^T e^{-\rho(\xi-s)}(u(C_\xi)-u(Y_\xi))d\xi \right)ds \right],
\end{aligned}
\end{eqnarray}
where $\lambda >0$ is the Lagrange multiplier associated with the promise-keeping constraint (\ref{eq:EN1}) at each time $t\geq 0$ and $e^{-r(s-t)}\eta_s \geq 0 $ is the Lagrange multiplier associated with the participation constraint (\ref{eq:EN2}) at each time $s \in [t,T]$.\\
Using integration by parts, the third term of right-hand side in (\ref{eq:LGR}) can be given as follows:
\begin{eqnarray}\label{eq:IB}
\begin{aligned}
&\mathbb{E}_t\left[\int_t^T e^{-r(s-t)}  \eta_s \left(\int_s^T e^{-\rho(\xi-s)}(u(C_\xi)-u(Y_\xi))d\xi \right)ds \right]\\
=&\mathbb{E}_t\left[\int_t^T e^{-\rho(s-t)} \cdot e^{(\rho-r)(s-t)}  \eta_s \left(\int_s^T e^{-\rho(\xi-s)}(u(C_\xi)-u(Y_\xi))d\xi \right)ds \right]\\
=&\mathbb{E}_t\left[\int_t^T  d\left(\int_t^s e^{(\rho-r)(\xi-t)}  \eta_\xi d\xi\right)\left(\int_s^T e^{-\rho(\xi-t)}(u(C_\xi)-u(Y_\xi))d\xi \right)ds \right]\\
=&\mathbb{E}_t\left[\int_t^T\left(\int_t^se^{(\rho-r)(\xi-t)}\eta_\xi d\xi\right)e^{-\rho(s-t)}(u(C_s)-u(Y_s))ds\right].
\end{aligned}
\end{eqnarray}
Plugging th {e} equation \eqref{eq:IB} into the Lagrangian \eqref{eq:LGR}, we obtain
\begin{eqnarray}\label{eq:LGR2}
\begin{aligned}
{\bf L} = & \mathbb{E}_t\left[ \int_t^T e^{-r(s-t)}  (Y_s-C_s)ds \right] +\lambda \mathbb{E}_t\left[\int_t^T e^{-\rho(s-t)}u(Y_s)ds\right]\\
+&\mathbb{E}_t\left[\int_t^T\left(\int_t^se^{(\rho-r)(\xi-t)}\eta_\xi d\xi+\lambda \right)e^{-\rho(s-t)}(u(C_s)-u(Y_s))ds\right]-\lambda w.
\end{aligned}
\end{eqnarray}
We define a \textit{costate process} $\{X_s\}_{s=t}^T$ as the cumulative amounts of the Lagrangian multipliers,
\begin{eqnarray}\label{eq:ND}
X_s \equiv \int_t^s e^{(\rho-r)(\xi-t)}\eta_\xi d\xi  + \lambda, \;\;\;\;X_t=\lambda.
\end{eqnarray} 
This process is non-decreasing, continuous, and satisfies
$$
dX_s=e^{(\rho-r)(s-t)}\eta_s ds,\;\;\;\;s\in[t,T].
$$
Then we can write down 
\begin{eqnarray}\label{eq:LGR3}
\begin{aligned}
{\bf L} =& \mathbb{E}_t\left[ \int_t^T e^{-r(s-t)}  (Y_s-C_s)ds \right] +\lambda \mathbb{E}_t\left[\int_t^T e^{-\rho(s-t)}u(Y_s)ds\right]\\
+&\mathbb{E}_t\left[\int_t^T X_s e^{-\rho(s-t)}(u(C_s)-u(Y_s))ds\right]-\lambda w.\\
\end{aligned}
\end{eqnarray}
For every enforceable consumption plan $\{C_s\}_{s=t}^{T}$, we can deduce that
\begin{eqnarray}
\mathbb{E}_t\left[\int_t^T e^{-r(s-t)}(Y_s-C_s)ds\right] \le \sup_{\{C_t\}}{\bf L}.
\end{eqnarray}
To derive the dual problem, we first choose the consumption at each time to obtain the maximum of the Lagrangian \eqref{eq:LGR3}, which takes the following form:
\begin{eqnarray}\label{eq:LGR4}
\begin{aligned}
{\bf L}(X) =& \sup_{\{C_t\}} \left\{ \mathbb{E}_t \left[\int_t^T e^{-r(s-t)}\Big(e^{-(\rho-r)(s-t)}X_su(C_s)-C_s\Big)ds\right]\right.\\ +&\left.\mathbb{E}_t\left[\int_t^T e^{-r(s-t)}Y_s-e^{-\rho(s-t)}X_su(Y_s) ds\right]+\lambda \mathbb{E}_t \left[\int_t^T e^{-\rho(s-t)}u(Y_s)ds \right]-\lambda w \right\}.\\
\end{aligned}
\end{eqnarray}      
By the first-order condition for a consumption plan $\{C_s\}_{s=t}^T$ in (\ref{eq:LGR4}), the optimal consumption $\{C^*_s\}_{s=t}^T$ is given as:
\begin{eqnarray}\label{eq:optimal_C}
C^*_s = (u^{\prime})^{-1}\left(\dfrac{e^{(\rho-r)(s-t)}}{X_s}\right)=(e^{-(\rho-r)(s-t)}X_s)^{\frac{1}{\gamma}},\;\;\;\;s\in[t,T].
\end{eqnarray}
Then {\bf L}$(X)$ can be given by
\begin{eqnarray}\label{eq:LGR5}
\begin{aligned}
{\bf L}(X) =& \mathbb{E}_t \left[\int_t^T e^{-r(s-t)} \Big( \tilde{u}(e^{-(\rho-r)(s-t)}X_s) +Y_s - e^{-(\rho-r)(s-t)}X_s u(Y_s)\Big)ds\right] \\
+&\lambda \mathbb{E}_t \left[\int_t^T e^{-\rho(s-t)}u(Y_s)ds \right]-\lambda w. \\
\end{aligned}
\end{eqnarray}
To ensure that the Lagrangian \eqref{eq:LGR5} is finite, we assume the following integrability conditions:
\begin{eqnarray}\label{eq:integrability}
\begin{split}
&{\mathbb{E}_t\left[\int_{t}^{T}e^{-\rho (s-t)}|u(Y_s)|X_s ds \right]<\infty,}\\
&\mathbb{E}_t\left[\int_{t}^{T}e^{-r (s-t)} |\tilde{u}(X_s e^{-(\rho-r)(s-t)})|ds\right]<\infty.
\end{split}
\end{eqnarray}
Let us define 
\begin{eqnarray}
\begin{split}\label{eq:fun_cal_J}
\mathcal{J}(t,\lambda,y,X)\equiv& \mathbb{E}_t\left[\int_t^T e^{-r(s-t)} \Big( \tilde{u}(e^{-(\rho-r)(s-t)}X_s) +Y_s - e^{-(\rho-r)(s-t)}X_s u(Y_s)\Big)ds  \right] \\
+&\lambda \mathbb{E}_t \left[\int_t^T e^{-\rho(s-t)}u(Y_s)ds \right].
\end{split}
\end{eqnarray}
{By the definition of the Lagrangian ${\bf L}(X)$ and the function $\mathcal{J}$} defined in \eqref{eq:fun_cal_J},
\begin{eqnarray*}
	\begin{split}
		\mathbb{E}_t\left[\int_t^T e^{-r(s-t)}(Y_s-C_s)ds\right]\le \mathcal{J}(t,\lambda,y,X)-\lambda w.
	\end{split}
\end{eqnarray*}
{Then, it is clear that}
\begin{eqnarray*}
	\mathbb{E}_t\left[\int_t^T e^{-r(s-t)}(Y_s-C_s)ds\right] \le \inf_{\lambda>0, X\in \mathcal{ND}(\lambda)}\left[\mathcal{J}(t,\lambda,y,X)-\lambda w\right]{,}
\end{eqnarray*}
where $\mathcal{ND}(\lambda)$ denotes the set of all positive non-decreasing, adaptable with respect to $\mathcal{F}$, right-continuous processes $X$ with left-limits (RCLL) and starting at $X_t=\lambda$, satisfying \eqref{eq:integrability}.

Hence, the value function $V(t,y,w)$, the maximized utility value of the principal,  satisfies the following inequality:
\begin{eqnarray}
\begin{split}\label{eq:du1}
V(t,y,w)=\sup_{C \in \Gamma_t(y,w)}\mathbb{E}_t\left[\int_t^T e^{-r(s-t)}(Y_s-C_s)ds\right]\le \inf_{\lambda>0, X\in \mathcal{ND}(\lambda)}\left[\mathcal{J}(t,\lambda,y,X)-\lambda w\right].
\end{split}
\end{eqnarray}
We will show in Theorem \ref{thm:main} that the maximized value is indeed equal to the right-hand side of the inequality in \eqref{eq:du1} with the infimum being replaced by the minimum, i.e.,
\begin{eqnarray}
\begin{split}\label{eq:du2}
V(t,y,w)=\min_{\lambda>0, X\in \mathcal{ND}(\lambda)}\left[\mathcal{J}(t,\lambda,y,X)-\lambda w\right]= \min_{\lambda>0}\left[\min_{X\in \mathcal{ND}(\lambda)}\mathcal{J}(t,\lambda,y,X)-\lambda w\right].
\end{split}
\end{eqnarray}
That is, we should choose the process $X$ to minimize ${\bf L}(X)$:
\begin{eqnarray}\label{eq:minimize_L}
\inf_{X\in \mathcal{ND}(\lambda)}{\bf L}(X).
\end{eqnarray}    
We now study the minimization problem inside the bracket of the right-hand side of the last equality in \eqref{eq:du2}, which we will call {as} the dual problem of Problem \ref{pr:main}.
\begin{pr}[Dual problem]\label{pr:dual}~
	Given $\lambda>0$, consider the minimization problem:
	\begin{eqnarray}\label{eq:dual_value}
	\begin{split}
	J(t,\lambda,y)=&\inf_{X \in \mathcal{ND}(\lambda)}\mathcal{J}(t,\lambda,y,X).
	\end{split}
	\end{eqnarray}
\end{pr}
To obtain the optimal costate process $\{X_s^*\}_{s=t}^T$, we transform Problem \ref{pr:dual} to the \textit{optimal stopping  problem}. This transformation is due to the existence of one-to-one correspondence between the set of all costate processes $\{X_s\}_{s=t}^T\in\mathcal{ND}(\lambda)$
and the set of all infinite series of $\mathcal{F}$-stopping times $\{\tau(x)\}_{x\ge\lambda}$ taking values in $[t,T]$ which is non-decreasing and left-continuous with right limits as a function of $x$.

The correspondence is given by
$$
\tau(x)=\inf\{s\ge t \mid X_s \ge x\}\wedge T.
$$
The problem of choosing a non-decreasing process $\{X_s\}_{s=t}^T$ is similar to an irreversible incremental investment problem studied by \citet{Pindyck88} and \citet{DixitPindyck}. They consider the capacity expansion decision of a firm as a series of optimal stopping problems: for each level of capacity, there is  {a corresponding} stopping problem where the firm chooses the optimal time to expand its capacity to  {an appropriate} level. Based on this idea, we transform Problem \ref{pr:dual} into a series of optimal stopping problems in the following lemma.
\begin{lem}\label{thm:lem1}~
	We can write the dual value function as follows:
	\begin{eqnarray}\label{eq:dual_ex1}
	J(t,\lambda,y)= -y^{1-\gamma}  \int_{\lambda}^{\infty} \left( \sup_{\tau(x)\in[t,T]}\mathbb{E}_t^{\mathbb{Q}} \left[ e^{-\hat{\rho}(\tau(x)-t)} h(\tau(x),x\mathcal{H}_{\tau(x)}) \right]\right)dx + J_0(t,\lambda,y){,}
	\end{eqnarray}
	where 
	\begin{eqnarray}
	\begin{aligned}
	\mathcal{H}_s &\equiv e^{-(\rho-r)(s-t)}Y_s^{-\gamma},\\
	h(t,z)&\equiv\frac{1}{1-\gamma}\left(\frac{1-e^{-\hat{\rho}(T-t)}}{\hat{\rho}}-\frac{1-e^{-K(T-t)}}{K}z^{\frac{1}{\gamma}-1}\right),\\
	{J}_0(t,\lambda,y)&\equiv\frac{\gamma}{1-\gamma}\frac{1-e^{-K(T-t)}}{K}\lambda^{\frac{1}{\gamma}}+\frac{1-e^{-\hat{r}(T-t)}}{\hat{r}}y .\\
	\end{aligned}
	\end{eqnarray}
	and the measure $\mathbb{Q}$ is defined in the proof. $\mathbb{E}^{\mathbb{Q}}[\cdot]$ is the expectation with respect to measure $\mathbb{Q}$. The corresponding standard Brownian motion $B_s^{\mathbb{Q}}$ is defined as $$ B_s^{\mathbb{Q}}\equiv B_s-(1-\gamma)\sigma s,\;s\in[t,T].$$
\end{lem}
\noindent{\bf Proof.} Define a function $f$ as follows
$$
f(x)=e^{-r(s-t)}\left(\tilde{u}(e^{-(\rho-r)(s-t)}x) +Y_s - e^{-(\rho-r)(s-t)}x \cdot u(Y_s) \right).
$$
By assigning (\ref{eq:CRRA}) and (\ref{eq:CRRA_D}) to $f$,
\begin{eqnarray}
f(x)=e^{-r(s-t)}\left(\frac{\gamma}{1-\gamma}(e^{-(\rho-r)(s-t)}x)^{\frac{1}{\gamma}} +Y_s - e^{-(\rho-r)(s-t)}x \frac{1}{1-\gamma}Y_s^{1-\gamma} \right)
\end{eqnarray}
and
\begin{eqnarray}\label{eq:fprime}
f^{\prime}(x)=\frac{e^{-\rho(s-t)}}{1-\gamma}\left((e^{-(\rho-r)(s-t)}x)^{\frac{1}{\gamma}-1}  -  Y_s^{1-\gamma} \right).
\end{eqnarray}
Then,
\begin{eqnarray}\label{eq:ST}
\begin{aligned}
&J(t,\lambda,y)\\
=& \inf_{X_s \in \mathcal{ND}(\lambda)} \mathbb{E}_t \left[\int_t^T f(X_s)ds \right]+\lambda\mathbb{E}_t\left[\int_t^T e^{-\rho(s-t)}u(Y_s)ds\right]\\
=&  \inf_{X_s \in \mathcal{ND}(\lambda)} \mathbb{E}_t \left[\int_t^T \left(\int_{X_t}^{X_s} f^{\prime}(x)dx+f(X_t)\right)ds\right]+\lambda\mathbb{E}_t\left[\int_t^T e^{-\rho(s-t)}u(Y_s)ds\right]\\
=&  \inf_{X_s \in \mathcal{ND}(\lambda)} \mathbb{E}_t \left[\int_t^T \left(\int_{\lambda}^{\infty} f^{\prime}(x)\cdot {\bf 1}_{\{x\leq X_s\}}dx\right)ds\right]+\mathbb{E}_t\left[\int_t^T f(\lambda) ds \right]+\lambda\mathbb{E}_t\left[\int_t^T e^{-\rho(s-t)}u(Y_s)ds\right]\\
=&  \inf_{X_s \in \mathcal{ND}(\lambda)} \mathbb{E}_t \left[\int_t^T \left(\int_{\lambda}^{\infty} f^{\prime}(x)\cdot {\bf 1}_{\{x\leq X_s\}}dx\right)ds\right]+\lambda\mathbb{E}_t\left[\int_t^T e^{-\rho(s-t)}u(Y_s)ds\right]\\
+&\mathbb{E}_t\left[\int_t^T e^{-r(s-t)}\left(\tilde{u}(e^{-(\rho-r)(s-t)}\lambda) +Y_s - e^{-(\rho-r)(s-t)}\lambda \cdot u(Y_s) \right) ds \right]\\
=&\inf_{X_s \in \mathcal{ND}(\lambda)} \mathbb{E}_t \left[\int_t^T \left(\int_{\lambda}^{\infty} f^{\prime}(x)\cdot {\bf 1}_{\{w\leq X_s\}}dx\right)ds\right] + \mathbb{E}_t\left[\int_t^T e^{-r(s-t)}\left(\tilde{u}(e^{-(\rho-r)(s-t)}\lambda) +Y_s \right) ds \right]\\
=&\inf_{\tau(x) \in [t,T]} \mathbb{E}_t \left[\int_{\lambda}^{\infty} \left(\int_{t}^{T} f^{\prime}(x)\cdot {\bf 1}_{\{s> \tau(x)\}}ds\right)dx\right] + \mathbb{E}_t\left[\int_t^T e^{-r(s-t)}\left(\tilde{u}(e^{-(\rho-r)(s-t)}\lambda) +Y_s \right) ds \right]\\
=&\int_{\lambda}^{\infty} \left(\inf_{\tau(x) \in [t,T]} \mathbb{E}_t \left[\int_{\tau(x)}^{T} f^{\prime}(x)ds\right]\right)dx + \mathbb{E}_t\left[\int_t^T e^{-r(s-t)}\left(\tilde{u}(e^{-(\rho-r)(s-t)}\lambda) +Y_s \right) ds \right].\\
\end{aligned}
\end{eqnarray}
where a stopping time $\tau(x)$ is defined by
\begin{eqnarray}\label{eq:ST1}
\tau(x) = \inf\left\{s\geq t | X_s \geq x\right\}\wedge T,\;\;\;\forall x\ge\lambda.
\end{eqnarray}
Note that Fubini's theorem implies that 
$$
\inf_{X_s \in \mathcal{ND}(\lambda)} \mathbb{E}_t \left[\int_t^T \left(\int_{\lambda}^{\infty} f^{\prime}(x)\cdot {\bf 1}_{\{x\leq X_s\}}dx\right)ds\right]=\inf_{\tau(x) \in [t,T]} \mathbb{E}_t \left[\int_{\lambda}^{\infty} \left(\int_{t}^{T} f^{\prime}(x)\cdot {\bf 1}_{\{s> \tau(x)\}}ds\right)dx\right].
$$
Define a exponential martingale process $Z_s^t$ and corresponding probability measure $\mathbb{Q}$ for each $s \in [t,T]$,
$$
Z_s^t=\exp\left\{-\frac{1}{2}(1-\gamma)^2\sigma^2(s-t) + (1-\gamma)\sigma (B_s-B_t) \right\},~~\textrm{and}~~~\frac{d\mathbb{Q}}{d\mathbb{P}}=Z_s^t,
$$
{respectively}. 

Girsanov's theorem implies that
$$
dB_s^{\mathbb{Q}}=dB_s - (1-\gamma)\sigma ds,\;\;\;\;s\in[t,T]
$$
is a standard Brownian motion under the measure $\mathbb{Q}$.

Since $Y_s = Y_t  e^{(\mu-\frac{1}{2}\sigma^2)(s-t)+\sigma(B_s-B_t)}$ and $Y_t=y$, let us define $Y_s^t \equiv e^{(\mu-\frac{1}{2}\sigma^2)(s-t)+\sigma(B_s-B_t)}$. By using (\ref{eq:fprime}), the first term of the last equation in (\ref{eq:ST}) can be derived as follows.
\begin{footnotesize}
	\begin{eqnarray}\label{eq:ST_1}
	\begin{aligned}
	&\inf_{\tau(x) \in [t,T]}\mathbb{E}_t\left[\int_{\tau(x)}^T f^{\prime}(x)ds\right]\\
	=&\inf_{\tau(x) \in [t,T]}\mathbb{E}_t\left[\int_{\tau(x)}^T \frac{e^{-\rho(s-t)}}{1-\gamma}\left((e^{-(\rho-r)(s-t)}x)^{\frac{1}{\gamma}-1}  -  Y_s^{1-\gamma} \right) ds \right]\\
	=&\inf_{\tau(x) \in [t,T]}\frac{1}{1-\gamma}\mathbb{E}_t \left[e^{-\rho(\tau(x)-t)}\mathbb{E}_{\tau(x)}\left[\int_{\tau(x)}^T e^{-\rho(s-\tau(x))}\left((e^{-(\rho-r)(s-t)}x)^{\frac{1}{\gamma}-1}-  Y_s^{1-\gamma}\right)ds \right]\right]\\
	=&\inf_{\tau(x) \in [t,T]}\frac{y^{1-\gamma}}{1-\gamma}\mathbb{E}_t \left[ e^{-\rho(\tau(x)-t)}  (Y^t_{\tau(x)})^{1-\gamma}\mathbb{E}_{\tau(x)}\left[\int_{\tau(x)}^T e^{-\rho(s-\tau(x))} (Y_s^{\tau(x)})^{1-\gamma} \left(\left(\frac{e^{-(\rho-r)(s-t)}x}{(Y_s)^{\gamma}}\right)^{\frac{1}{\gamma}-1}-  1\right)ds \right]\right]\\
	=&\inf_{\tau(x) \in [t,T]}\frac{y^{1-\gamma}}{1-\gamma}\mathbb{E}_t \left[ e^{-\hat{\rho}(\tau(x)-t)}  Z^t_{\tau(x)}\mathbb{E}_{\tau(x)}\left[\int_{\tau(x)}^T e^{-\hat{\rho}(s-\tau(x))}Z_s^{\tau(x)}\left(\left(\frac{e^{-(\rho-r)(s-t)}x}{(Y_s)^{\gamma}}\right)^{\frac{1}{\gamma}-1}-  1\right)ds \right]\right]\\
	=&\inf_{\tau(x) \in [t,T]}\frac{y^{1-\gamma}}{1-\gamma}\mathbb{E}_t^{\mathbb{Q}} \left[ e^{-\hat{\rho}(\tau(x)-t)} \mathbb{E}_{\tau(x)}^{\mathbb{Q}}\left[\int_{\tau(x)}^T e^{-\hat{\rho}(s-\tau(x))}\left(\left(\frac{e^{-(\rho-r)(s-t)}x}{(Y_s)^{\gamma}}\right)^{\frac{1}{\gamma}-1}-  1\right)ds \right]\right]\\
	=&\inf_{\tau(x) \in [t,T]}\frac{y^{1-\gamma}}{1-\gamma}\mathbb{E}_t^{\mathbb{Q}}  \left[ e^{-\hat{\rho}(\tau(x)-t)} \mathbb{E}_{\tau(x)}^{\mathbb{Q}}\left[\int_{\tau(x)}^T\left( e^{-\hat{\rho}(s-\tau(x))} \left(x\mathcal{H}_{s}\right)^{\frac{1}{\gamma}-1}\cdot e^{-(\rho-r)(s-\tau(x))\left(\frac{1}{\gamma}-1\right)}(Y_s^{\tau(x)})^{\gamma-1}\right.\right.\right.\\ -&\left.\left.\left.e^{-\hat{\rho}(s-\tau(x))} \right)ds   \right]\right]\\
	=& \inf_{\tau(x) \in [t,T]}\frac{y^{1-\gamma}}{1-\gamma}\mathbb{E}_t^{\mathbb{Q}} \left[ e^{-\hat{\rho}(\tau(x)-t)}\left(\frac{1-e^{-K(T-\tau(x))}}{K}(x\mathcal{H}_{\tau(x)})^{\frac{1}{\gamma}-1} - \frac{1-e^{-\hat{\rho}(T-\tau(x))}}{\hat{\rho}} \right) \right].
	\end{aligned}
	\end{eqnarray}	
\end{footnotesize}
\begin{rem}\label{thm:Qprocess}~ Under the measure $\mathbb{Q}$,
	\begin{eqnarray}\label{eq:Hprocess}
	\begin{aligned}
	&dY_s=(\mu+(1-\gamma)\sigma^2)Y_sds + \sigma Y_s dB_s^{\mathbb{Q}},\\
	&d\mathcal{H}_s=(\hat{r}-\hat{\rho}+\sigma^2\gamma^2)\mathcal{H}_sds-\gamma\sigma \mathcal{H}_sdB_s^{\mathbb{Q}}.
	\end{aligned}
	\end{eqnarray}
	The process $\mathcal{H}_s$ can be easily derived by It{\^o}'s lemma. We will use this process to prove variational inequality(VI) later.
\end{rem}
The other term of last equation in (\ref{eq:ST}) can be directly derived.
\begin{eqnarray}\label{eq:ST_2}
\begin{aligned}
&\mathbb{E}_t\left[\int_t^T e^{-r(s-t)}\left(\tilde{u}(e^{-(\rho-r)(s-t)}\lambda) +Y_s \right) ds \right]=\frac{\gamma}{1-\gamma} \frac{1-e^{-K(T-t)}}{K}\lambda^{\frac{1}{\gamma}}+\frac{1-e^{-\hat{r}(T-t)}}{\hat{r}}y.
\end{aligned}
\end{eqnarray} \\
By (\ref{eq:ST_1}) and (\ref{eq:ST_2}),
\begin{footnotesize}
	\begin{eqnarray}
	\begin{aligned}
	&J(t,\lambda,y)\\
	=& \int_{\lambda}^{\infty} \left( \inf_{\tau(x) \in [t,T]}\frac{y^{1-\gamma}}{1-\gamma}\mathbb{E}_t^{\mathbb{Q}} \left[ e^{-\hat{\rho}(\tau(x)-t)}\left(\frac{1-e^{-K(T-\tau(x))}}{K}(x\mathcal{H}_{\tau(x)})^{\frac{1}{\gamma}-1} - \frac{1-e^{-\hat{\rho}(T-\tau(x))}}{\hat{\rho}} \right) \right]\right)dx\\
	+&\frac{\gamma}{1-\gamma}\left(\frac{1-e^{-K(T-t)}}{K}\lambda^{\frac{1}{\gamma}}+\frac{1-e^{-\hat{r}(T-t)}}{\hat{r}}y \right)\\
	=&-y^{1-\gamma}  \int_{\lambda}^{\infty} \left( \sup_{\tau(x) \in [t,T]}\mathbb{E}_t^{\mathbb{Q}} \left[ e^{-\hat{\rho}(\tau(x)-t)}\frac{1}{1-\gamma}\left(\frac{1-e^{-\hat{\rho}(T-\tau(x))}}{\hat{\rho}}-\frac{1-e^{-K(T-\tau(x))}}{K}(x\mathcal{H}_{\tau(x)})^{\frac{1}{\gamma}-1} \right) \right]\right)dx\\
	+&{ \frac{\gamma}{1-\gamma}\left(\frac{1-e^{-K(T-t)}}{K}\lambda^{\frac{1}{\gamma}}+\frac{1-e^{-\hat{r}(T-t)}}{\hat{r}}y \right)}.\\
	\end{aligned}
	\end{eqnarray}	
\end{footnotesize}
This ends the proof of Lemma \ref{thm:lem1}.
\hfill$\Box$\medskip

In Lemma \ref{thm:lem1}, the dual problem can be solved by converting it to the infinite series of optimal stopping time problems. Since the underlying process $(\mathcal{H}_s)_{s=t}^{T}$ has  {an} exponential form, however, the multiplicative factor $x$ can be absorbed into the initial condition problem. Thus, the problems can be  {combined}  {in}to a single problem{,} as shown below.
\begin{pr}[Optimal stopping problem]\label{pr:OS}~
	We consider the following optimal stopping problem: 
	$$
	g(t,z)=\sup_{\tau \in \mathcal{S}(t,T)} \mathbb{E}^{\mathbb{Q}}\left[e^{-\hat{\rho}(\tau-t)}h(\tau,\mathcal{H}_{\tau}) \bigm| \mathcal{H}_t=z\right]{,}
	$$
	where $\mathcal{S}(t,T)$ denotes the set of all stopping times of the filtration $\mathcal{F}$ taking values in $[t,T]$, and  
	$$
	h(t,z)=\frac{1}{1-\gamma}\left(\frac{1-e^{-\hat{\rho}(T-t)}}{\hat{\rho}}-\frac{1-e^{-K(T-t)}}{K}z^{\frac{1}{\gamma}-1}\right)
	$$
	and $(\mathcal{H}_s)_{s\geq t}$ satisfies the following stochastic diffusion process:
	$$
	d\mathcal{H}_s=(\hat{r}-\hat{\rho}+\sigma^2\gamma^2)\mathcal{H}_sds-\gamma\sigma \mathcal{H}_sdB_s^{\mathbb{Q}}.
	$$	
\end{pr}
Notice that Problem \ref{pr:OS} is equivalent to that of finding the optimal exercise time of an American option written on the underlying process $\{\mathcal{H}_s\}_{s=t}^{T}$ with payoff equal to $h(\tau,\mathcal{H}_\tau)$ at the time of exercise time $\tau$. The exercise time is characterized as the first time for the underlying process to hit the early exercise boundary(or free boundary), and thus the problem is to derive the early exercise boundary. \citet{K}, \citet{CJM}, and \citet{Det} provide an integral equation representation of the American option value from which one can derive a functional equation for the early exercise boundary. 

According to the standard technique  {for} the optimal stopping problem, $g(T,z)$ can be derived from the following variational inequality(VI). (See Chapter 2 of \citet{KS} or \citet{YK}):
\begin{eqnarray}\label{eq:VI1}
\begin{split}
\left\{
\begin{array}{l}
-\partial_t g-{\cal L} g=0,~~
\mbox{if}~~g(t,z)>h(t,z)~~\textrm{and}~~(t,z)\in \mathcal{M}_T,
\vspace{2mm} \\
-\partial_t g-{\cal L} g\geq 0,~~
\mbox{if}~~g(t,z)=h(t,z) ~~\textrm{and}~~(t,z)\in \mathcal{M}_T,
\vspace{2mm} \\
g(T,z)=h(T,z),\quad \forall \;z\in(0,+\infty){,}
\end{array}
\right.
\end{split}
\end{eqnarray}
where $\mathcal{M}_T=[0,T)\times(0,+\infty)$ and the operator $\mathcal{L}$ is generated by the process $\mathcal{H}_s$:
$$
\mathcal{L}= \frac{\gamma^2\sigma^2}{2}z^2 \partial_{zz}+(\hat{r}-\hat{\rho}+\sigma^2\gamma^2)z\partial_z-\hat{\rho}.
$$

\section{Analysis of variational inequality arising from Problem \ref{pr:OS}}\label{sec:B}

{In this section we provide a complete self-contained derivation of the solution to VI \eqref{eq:VI1} arising from Problem \ref{pr:OS} by borrowing the ideas and proofs in \citet{YK}.}

For convenience of proof, we substitute the above VI \eqref{eq:VI1} for the following to make the lower obstacle 0.
$$
Q(t,z)=g(t,z)-h(t,z).
$$
Then, the (\ref{eq:VI1}) can be converted to
\begin{eqnarray}\label{eq:VI2}
\begin{split}
\left\{
\begin{array}{l}
-\partial_t Q-{\cal L} Q=\dfrac{1}{1-\gamma}\left(z^{\frac{1}{\gamma}-1}-1\right),~~
\mbox{if}~~Q(t,z)>0~~\textrm{and}~~(t,z)\in \mathcal{M}_T,
\vspace{2mm} \\
-\partial_t Q-{\cal L} Q\geq \dfrac{1}{1-\gamma}\left(z^{\frac{1}{\gamma}-1}-1\right),~~
\mbox{if}~~Q(t,z)=0 ~~\textrm{and}~~(t,z)\in \mathcal{M}_T,
\vspace{2mm} \\
Q(T,z)=0,\quad \forall \;z\in(0,+\infty).
\end{array}
\right.
\end{split}
\end{eqnarray}
Now, we will prove the existence and uniqueness of $W^{1,2}_{p,loc}$ solution {$(p\geq1)$} to VI \eqref{eq:VI2} and describe properties of the solution.
\begin{lem}\label{thm:lemQ}~
	VI (\ref{eq:VI2}) has a unique strong solution $Q$ satisfying the following properties:\\
	\noindent 1. $Q \in W^{1,2}_{p,loc}(\mathcal{M}_T) \cap C(\widetilde{\mathcal{M_T}})$ for any $p \geq 1 $ and $\partial_z Q \in C(\widetilde{\mathcal{M}_T})$, where $\widetilde{\mathcal{M}}_T=[0,T]\times (0,+\infty)$.\\
	\noindent 2. $\partial_zQ \geq 0$ a.e. in $\widetilde{\mathcal{M}_T}$ and $\partial_t Q \leq 0$ a.e. in $\widetilde{\mathcal{M}_T}$.
\end{lem}
\noindent{\bf Proof.} 
1. To use the Theorem of \cite{F2}, replace the PDE operator $\mathcal{L}$ with a non-degenerate parabolic equation. 

Define
$$
\z=\log z,~~~~ \bar{Q}(t,\z)=Q(t,z).
$$
Then $\bar{Q}(t,\z)$ satisfies 
\begin{eqnarray}\label{eq:VI3}
\begin{split}
\left\{
\begin{array}{l}
-\partial_t \bar{Q}-\bar{{\cal L}} \bar{Q}=\dfrac{1}{1-\gamma}\left(e^{\z(\frac{1}{\gamma}-1)}-1\right),~~
\mbox{if}~~\bar{Q}(t,\z)>0~~\textrm{and}~~{(t,e^\z)\in \mathcal{M}_T}
\vspace{2mm} \\
-\partial_t \bar{Q}-\bar{{\cal L}} \bar{Q}\geq \dfrac{1}{1-\gamma}\left(e^{\z(\frac{1}{\gamma}-1)}-1\right),~~
\mbox{if}~~\bar{Q}(t,\z)=0 ~~\textrm{and}~~{(t,e^\z)\in \mathcal{M}_T}
\vspace{2mm} \\
\bar{Q}(T,\z)=0,\quad \forall \;\z\in(0,+\infty).
\end{array}
\right.
\end{split}
\end{eqnarray}
where
$$
\bar{{\cal L}}= \frac{\gamma^2\sigma^2}{2} \partial_{\z\z}+(\hat{r}-\hat{\rho}+\frac{\sigma^2\gamma^2}{2})\partial_\z-\hat{\rho}.
$$
Since the inhomogeneous term `$\dfrac{1}{1-\gamma}\left(e^{\z(\frac{1}{\gamma}-1)}-1\right)$', the lower obstacle `0' and the terminal value `0' are all smooth functions, we can easily show that VI(\ref{eq:VI3}) has a unique solution satisfying $\bar{Q} \in W^{1,2}_{p,loc}(\mathcal{M}_T) \cap C(\widetilde{\mathcal{M_T}})$ for any $p \geq 1 $ and $\partial_\z \bar{Q} \in C(\widetilde{\mathcal{M}_T})$(See \cite{F2}).\\

2. Let us denote $\widetilde{Q}(t,z)=Q(t,\eta z)$ for any $\eta >1$. Then $\widetilde{Q}$ satisfies following Variational Inequality:
\begin{eqnarray}\label{eq:VI4}
\begin{split}
\left\{
\begin{array}{l}
-\partial_t \widetilde{Q}-{\cal L} \widetilde{Q}=\dfrac{1}{1-\gamma}\left((\eta z)^{\frac{1}{\gamma}-1}-1\right),~~
\mbox{if}~~\widetilde{Q}(t,z)>0~~\textrm{and}~~(t,z)\in \mathcal{M}_T,
\vspace{2mm} \\
-\partial_t \widetilde{Q}-{\cal L} \widetilde{Q}\geq \dfrac{1}{1-\gamma}\left((\eta z)^{\frac{1}{\gamma}-1}-1\right),~~
\mbox{if}~~\widetilde{Q}(t,z)=0 ~~\textrm{and}~~(t,z)\in \mathcal{M}_T,
\vspace{2mm} \\
\widetilde{Q}(T,z)=0\quad \forall \;z\in(0,+\infty).
\end{array}
\right.
\end{split}
\end{eqnarray}
For any $\eta >1$, we can easily check that the inhomogeneous term of $\widetilde{Q}$ is greater than $Q$: 
$$
\frac{1}{1-\gamma}\left((\eta z)^{\frac{1}{\gamma}-1}-1\right) > \frac{1}{1-\gamma}\left(z^{\frac{1}{\gamma}-1}-1\right),~~~~~~\textrm{for}~\forall~\gamma>0 (\gamma \neq 1).
$$
In addition, since the terminal value of $Q$ and $\widetilde{Q}$ are the same, the comparison theory for VI implies that $\widetilde{Q}(t,z)=Q(t,\eta z) \geq Q(t,z)$ for any $\eta >1$ and $(t,z) \in \mathcal{M}_T$. So we obtain $\partial_z Q \geq 0$ in $\mathcal{M}_T$.\\
Define $\hat{Q}(t,z)=Q(t-\delta,z)$ with $\delta>0$ being sufficiently small. Then $\hat{Q}$ follows:
\begin{eqnarray}\label{eq:VI5}
\begin{split}
\left\{
\begin{array}{l}
-\partial_t \hat{Q}-{\cal L} \hat{Q}=\dfrac{1}{1-\gamma}\left(z^{\frac{1}{\gamma}-1}-1\right),~~
\mbox{if}~~\hat{Q}(t,z)>0~~\textrm{and}~~(t,z)\in \mathcal{M}_T,
\vspace{2mm} \\
-\partial_t \hat{Q}-{\cal L} \hat{Q}\geq \dfrac{1}{1-\gamma}\left(z^{\frac{1}{\gamma}-1}-1\right),~~
\mbox{if}~~\hat{Q}(t,z)=0 ~~\textrm{and}~~(t,z)\in \mathcal{M}_T,
\vspace{2mm} \\
\hat{Q}(T,z)=0\quad \forall \;z\in(0,+\infty).
\end{array}
\right.
\end{split}
\end{eqnarray}
Since $\hat{Q}(T,z)(=Q(T-\delta,z)) \geq Q(T,z)$ for any $\delta>0,z>0$, the following can be deduced by the comparison theory for the same reason: $\hat{Q}(t,z)=Q(t-\delta,z) \leq Q(t,z)$. Therefore we can prove that $\partial_t Q \leq 0$, a.e. 
\hfill$\Box$\medskip

We can define two regions derived from the Variational Inequality (\ref{eq:VI2}):
\begin{eqnarray}\label{eq:def_region1}
\Omega_1=\left\{(t,z)~ |~ Q(t,z)=0 \right\},~~~\Omega_2=\left\{(t,z)~ |~ Q(t,z)>0 \right\}.
\end{eqnarray}
Since $\partial_z Q \ge 0$, we can define the free boundary $z^{\star}(t)$ as follows:
\begin{eqnarray}
z^{\star}(t)=\inf \left\{ z\geq 0 ~|~ Q(t,z)>0 \right\},~~~~\forall t\in[0,T).
\end{eqnarray}
The two regions can be defined according to the boundary as follows. 
\begin{eqnarray}
\begin{aligned}\label{eq:def_region2}
&\Omega_1=\left\{(t,z)~ |~ 0<z\leq z^{\star}(t),~ t \in [0,T] \right\},\\
&\Omega_2=\left\{(t,z)~ |~ z> z^{\star}(t),~ t \in [0,T] \right\}.
\end{aligned}
\end{eqnarray}
\begin{lem}\label{lem-cinfinity}~The free boundary $z^{\star}$ is smooth, i.e., $z^{\star}(t)\in C[0,T]\cap C^{\infty}([0,T))$. Moreover, the solution $Q\equiv 0$ in $\Omega_1$,\;and $Q\in C^{\infty}\left(\{(t,z)\mid z \ge z^{\star}(t),\;t\in[0,T]\}\right)$, and $\partial_{t} Q \in C(\widetilde{\mathcal{M}}_{T})$.
\end{lem}
\noindent{\bf Proof.} 
By Lemma~\ref{thm:lemQ}, we obtain $\partial_{t} Q \le 0$ \rm{a.e.} in $\widetilde{\mathcal{M}}_{T}$. Moreover, the coefficient functions in the operator $\mathcal{L}$, the lower obstacle function, the terminal function, and the non-homogeneous term $\frac{1}{1-\gamma}\left(z^{\frac{1}{\gamma}-1}-1\right)$ are all smooth. Therefore, the regularity results in the lemma follow from Theorem 3.1 in ~\citet{F2}.
\hfill$\Box$\medskip

Consider the following function $Q_{\infty}$ as follow:
\begin{eqnarray}\label{eq:INFQ}
\begin{split}
Q_{\infty}(t,z)=
\left\{
\begin{array}{l}
\left(-\dfrac{1}{\gamma}\dfrac{1}{K}\dfrac{1}{\alpha_-}(z_{\infty})^{\frac{1}{\gamma}-1-\alpha_-}\right)z^{\alpha_-}+\dfrac{1}{1-\gamma}\left(\dfrac{1}{K}z^{\frac{1}{\gamma}-1}-\dfrac{1}{\hat{\rho}}\right),~~
\mbox{if}~~(t,z)\in\Omega_2^{\infty},~~
\vspace{2mm} \\
0,\;\;\;\;\;\;\;\;\;\;\;~~~~~~~~~~~~~~~~~~~~~~~~~~~~~~~~~~~~~~~~~~~~~~~~~~~~~~~~~~~~~~~~
\mbox{if}~~(t,z)\in\Omega_1^{\infty}, 
\vspace{2mm} 
\end{array}
\right.
\end{split}
\end{eqnarray}
where
\begin{eqnarray}\label{eq:INFZ}
\begin{aligned}
&\Omega_1^{\infty}=\left\{(t,z)~ |~ 0<z\leq z_{\infty},~ t \in [0,T] \right\},~~\Omega_2^{\infty}=\left\{(t,z)~ |~ z> z_{\infty},~ t \in [0,T] \right\}
\end{aligned}
\end{eqnarray}
and 
\begin{eqnarray}
z_{\infty}=\left(\frac{\hat{\rho}(\alpha_- \gamma + \gamma -1)}{K \alpha_- \gamma}\right)^{\frac{\gamma}{\gamma-1}}.
\end{eqnarray}
$\alpha_+$ and $\alpha_-$ are the positive and negative root of the following quadratic equation $f(\alpha)=0$, respectively:
\begin{eqnarray}\label{eq:f1}
f(\alpha)=\frac{\gamma^2 \sigma^2}{2}\alpha^2+(\hat{r}-\hat{\rho}+\frac{\gamma^2\sigma^2}{2})\alpha-\hat{\rho}=0.
\end{eqnarray}
It is easy to confirm that 
$$
f(-1)=-\hat{r}=\mu-r<0,\;\;\mbox{and}\;\alpha_- < -1.
$$
\begin{lem}\label{thm:lem2}~The solution $Q$ satisfy the following statements:\\
	\noindent 1.~$Q(t,z)\equiv0 $ in  $\Omega_1^\infty$.\\
	\noindent 2. $Q(t,z)>0$ in $[0,T) \times (z^T,\infty)$, where $z^T=1$.
\end{lem}
\noindent{\bf Proof.}\\
{1.} 
It is easy to check that
$$
Q_{\infty} \in  W^{1,2}_{p,loc}(\mathcal{M}_T) \cap C(\widetilde{\mathcal{M_T}})~~\textrm{for}~~p\geq 1 .
$$
Since $\alpha_-<-1$, we deduce
$$
\frac{\partial Q_{\infty}}{\partial z}=\frac{1}{\gamma K}z^{\frac{1}{\gamma}-2}\left(1-\left(\frac{z}{z_\infty}\right)^{\alpha_-+1-\frac{1}{\gamma}}\right)>0,~~\textrm{in}~~\Omega_2^{\infty}=\left\{(t,z)~ |~ z> z_{\infty},~ t \in [0,T] \right\}.
$$
We can see that $Q_{\infty}$ satisfies the following equality.
\begin{eqnarray}\label{eq:VI_infty}
\begin{split}
-\partial_t Q_{\infty}-{\cal L} Q_{\infty}=
\left\{
\begin{array}{l}
\dfrac{1}{1-\gamma}\left(z^{\frac{1}{\gamma}-1}-1\right),~~
\mbox{in}~~\Omega_2^{\infty}=\left\{(t,z)~ |~ z> z_{\infty},~ t \in [0,T] \right\},
\vspace{2mm} \\
0,\;\;\;~~~~~~~~~~~~~~~~~~~~~~
\mbox{in}~~\Omega_1^{\infty}=\left\{(t,z)~ |~ 0<z\leq z_{\infty},~ t \in [0,T] \right\}.
\vspace{2mm} \\
\end{array}
\right.
\end{split}
\end{eqnarray}
Prior to proving  $1.$ of Lemma \ref{thm:lem2} , we can first show the following inequality:
\begin{eqnarray}\label{eq:Z1}
z_\infty < z^T ~~~~\Longleftrightarrow~~~~\left(\frac{\hat{\rho}(\alpha_- \gamma + \gamma -1)}{K \alpha_- \gamma}\right)^{\frac{\gamma}{\gamma-1}}<1.
\end{eqnarray}
If $0<\gamma<1$, the above inequality is equivalent to
$$
\hat{\rho}(\alpha_-\gamma+\gamma-1) < K\alpha_-\gamma ~~~~\Longleftrightarrow~~~~(\hat{\rho}-K)\gamma \alpha_- < \hat{\rho}(1-\gamma).
$$
By Assumption \ref{thm:as1}, 
$$
\hat{\rho}-K=(\rho-r)\left(1-\frac{1}{\gamma} \right)-(1-\gamma)\mu +\gamma(1-\gamma)\frac{\sigma^2}{2} <0.
$$
It is enough to show 
$$
\alpha_->\frac{\hat{\rho}(1-\gamma)}{\gamma(\hat{\rho}-K)}.
$$
Since $\alpha_-$ is negative root of quadratic equation $f(\alpha)=0$ in \eqref{eq:f1}, we need to show $f\left(\frac{\hat{\rho}(1-\gamma)}{\gamma(\hat{\rho}-K)}\right)>0$. By simple calculations,
$$
f\left(\frac{\hat{\rho}(1-\gamma)}{\gamma(\hat{\rho}-K)}\right)=\frac{K \hat{\rho} \sigma^2 (1-\gamma)^2}{2\gamma^2(\hat{\rho}-K)^2}>0.
$$
The following $\gamma>1$ case is also the same as the previously proven case $0<\gamma<1$ and thus is omitted here.

By using (\ref{eq:VI_infty}) and the inequality $z_\infty < z^T$ in (\ref{eq:Z1}), 
$$
-\partial_t Q_{\infty}-{\cal L} Q_{\infty} \geq \frac{1}{1-\gamma}\left(z^{\frac{1}{\gamma}-1}-1\right)~~~\forall ~(t,z) \in \mathcal{M}_T
$$
Note that $\frac{\partial Q_{\infty}}{\partial z}(t,z_\infty)=Q_{\infty}(t,z_\infty)=0$ and $\frac{\partial Q_{\infty}}{\partial z}(t,z)>0$ in $\Omega_2$. Therefore $Q_{\infty}(t,z)>0$ in $\Omega_2$ and $Q_{\infty}$ satisfies the following VI:
\begin{eqnarray}\label{eq:VI_infty2}
\begin{split}
\left\{
\begin{array}{l}
-\partial_t Q_{\infty}-{\cal L} Q_{\infty}=\dfrac{1}{1-\gamma}\left(z^{\frac{1}{\gamma}-1}-1\right),~~
\mbox{in}~~Q_{\infty}>0~~\textrm{and}~~(t,z)\in \mathcal{M}_T,
\vspace{2mm} \\
-\partial_t Q_{\infty}-{\cal L} Q_{\infty} \geq \dfrac{1}{1-\gamma}\left(z^{\frac{1}{\gamma}-1}-1\right),~~
\mbox{in}~~Q_{\infty}=0~~\textrm{and}~~(t,z)\in \mathcal{M}_T,
\vspace{2mm} \\
Q_{\infty}(t,z)\geq 0,~~~~\forall~~(t,z)\in \mathcal{M}_T.
\end{array}
\right.
\end{split}
\end{eqnarray}
By the comparison principle for VI, we have
$$
Q(t,z)\leq Q_{\infty}(t,z),~~~~~\forall~ (t,z)~~ \textrm{in}~~ \mathcal{M}_T.
$$
So we can get the first result of this lemma $Q=0$~~in~$\Omega_1^\infty$.
$$
0\leq Q(t,z) \leq Q_{\infty}(t,z)\leq0,~~\textrm{in}~~\Omega_1^\infty=\left\{(t,z)~ |~ 0<z\leq z_{\infty},~ t \in [0,T] \right\}. 
$$
Furthermore, this means that 
$$
\Omega_1=\left\{(t,z)~ |~ 0<z\leq z^\star(t),~ t \in [0,T] \right\}~~\supseteq~~\Omega_1^\infty=\left\{(t,z)~ |~ 0<z\leq z_{\infty},~ t \in [0,T] \right\}.
$$
Thus, we obtain that $z^{\star}(t)\geq z_\infty$ in $[0,T]$.\\

\noindent {2.}	
Finally, let us prove the second property of this Lemma. VI \eqref{eq:VI2} implies that 
$$
Q=0,~~~ -\partial_t Q-{\cal L} Q-\frac{1}{1-\gamma}\left(z^{\frac{1}{\gamma}-1}-1\right) \geq 0,~~~~\textrm{in}~~\Omega_1.
$$
This leads to $\Omega_1 \subseteq [0,T]\times (0,z^T]$. Hence, it is obvious that $\Omega_2=\left\{(t,z)~ |~ Q(t,z)>0 \right\} \supseteq [0,T)\times(z^T,\infty)$. Thus, $Q(t,z)>0$ in $[0,T) \times (z^T,\infty)$. Moreover,  
$$
z^{\star}(t)\leq z^T(=1)~~\textrm{in}~~[0,T].
$$
This is the end of the lemma.
\hfill$\Box$\medskip

\begin{lem}\label{thm:FB}~
	The free boundary $z^\star(t)$, $t \in [0,T]$ is strictly increasing with the terminal point $z^{\star}(T)=\displaystyle\lim_{t\rightarrow T^-}z^{\star}(t)=z^T$. And $z_{\infty}<z^{\star}(t)<z^T$, ~~$\forall t \in [0,T)$.
\end{lem}
\noindent{\bf Proof.} 
By Lemma \ref{thm:lemQ}, 
$$
\partial_z Q \geq 0,~\partial_t Q \leq 0 ~~\textrm{a.e. in}~~\mathcal{M}_T,~~~~Q\in C(\widetilde{\mathcal{M}}_T).
$$
For any fixed $t_0 \in [0,T)$, 
$$
0\leq Q(t,z) \leq Q(t,z^{\star}(t_0)) \leq Q(t_0,z^{\star}(t_0))=0,~~~~~\forall t \in (t_0,T]~~ \textrm{and}~~ \forall z \in (0,z^\star(t_0)].
$$
By the definition of the free boundary $z^{\star}(t)$, 
$$
z^{\star}(t) \geq z^{\star}(t_0),~~~~0\leq t_0 \leq t <T.
$$
The above {inequality} shows that $z^\star(t)$ is just increasing function with respect to $t$. (We still need to show the strict increasing property of $z^\star(t)$).\\
Since $z^\star(t)$ is increasing,  the $z^\star(T)=\displaystyle\lim_{t\rightarrow T-}z^\star(t)$ exists. We know that $z^\star(t)\leq z^T$ in $\forall t \in[0,T]$. It is therefore sufficient to show that $z^\star(T)\geq z^T$. Otherwise, there exists interval $(z^\star(T),z^T)$ such that $[0,T) \times (z^\star(T),z^T) \subset \Omega_2$. So, we have shown that
\begin{eqnarray*}
	\begin{split}
		\left\{
		\begin{array}{l}
			-\partial_t Q-{\cal L} Q=\dfrac{1}{1-\gamma}\left(z^{\frac{1}{\gamma}-1}-1\right),~~
			\mbox{in}~~\Omega_2
			\vspace{2mm} \\
			Q(T,z)= 0,~~~~\forall~z \geq z^\star(T),~~~~~~Q(t,z^\star(t))=0~~~\forall~t\in[0,T].
		\end{array}
		\right.
	\end{split}
\end{eqnarray*}
By using the above fact, we can deduce that at time $T$
$$
\partial_t Q(T,z)=-{\cal L} Q(T,z)-\frac{1}{1-\gamma}\left(z^{\frac{1}{\gamma}-1}-1\right)=-\frac{1}{1-\gamma}\left(z^{\frac{1}{\gamma}-1}-1\right)>0~~~~\forall~z\in(z^\star(T),z^T).
$$
However, it is easy to see that the above inequality is inconsistent with the results of Lemma \ref{thm:lemQ}.

Finally, we show that $z^\star(t)$ is strictly increasing. Otherwise, there exists $t_1,t_2$ and $z_1$ such that $z^\star(t)=z_1$ for all $t \in[t_2,t_1]$ where $0\leq t_2 < t_1 \leq T$ and $z_1 \in [z_{\infty},z^T]$. Then, it is clear that $Q(t,z)=0$ for all $(t,z)\in[t_2,t_1]\times(0,z_1)$. Because of the continuity at the free boundary of $\partial_zQ$, $\partial_z Q (t,z_1)=0$ for all $t\in[t_2,t_1]$. Therefore, we obtain that  $\partial_tQ(t,z_1)=\partial_z\partial_zQ(t,z_1)=0$ for all $t\in[t_2,t_1]$.  In domain $[t_2,t_1)\times(z_1,\infty)$, $\partial_tQ$ satisfies
\begin{eqnarray*}
	\begin{split}
		\left\{
		\begin{array}{l}
			-\partial_t\partial_t Q-{\cal L} \partial_t Q=0,~\partial_t Q \leq 0,~~
			\mbox{in}~~[t_2,t_1)\times(z_1,\infty),
			\vspace{2mm} \\
			\partial_tQ(t,z_1)=0,~~~~\forall~t\in(t_2,t_1).
		\end{array}
		\right.
	\end{split}
\end{eqnarray*}
According to Hopf's boundary point lemma (See \cite{Liu}), we obtain that $\partial_z(\partial_tQ)<0$, which contradicts the $\partial_z\partial_tQ(t,z_1)=0$ in $t\in[t_2,t_1]$. So, the free boundary $z^{\star}(t)$ is strictly increasing. Thus, We conclude that $z_{\infty}<z^{\star}(t)<z^T$ for all $t\in[0,T)$
\hfill$\Box$\medskip
\begin{lem}\label{thm:estimate}
	~For $(t,z)\in\mathcal{M}_T$, 
	\begin{eqnarray}
	0\le \partial_z Q(t,z) \le \dfrac{1}{\gamma}\cdot\dfrac{1-e^{-K(T-t)}}{K}z^{\frac{1}{\gamma}-2}.
	\end{eqnarray}
\end{lem}
\noindent{\bf Proof.} 
From VI \eqref{eq:VI2}, $Q(t,z)$ satisfies 
\begin{eqnarray*}
	\begin{split}
		\begin{cases}
			&-\partial_t Q -\mathcal{L} Q =\dfrac{1}{1-\gamma}\left(z^{\frac{1}{\gamma}-1}-1\right),\;\;\;\mbox{in}\;\Omega_2,\\
			&Q(T,z)=0,\;\;\forall\;z\ge z^{\star}(T);\qquad Q(t,z^{\star}(t))=0,\;\;\forall\;t\in[0,T].
		\end{cases}
	\end{split}
\end{eqnarray*}
Since $\mathcal{L}(z\partial_z Q)=z\partial_z(\mathcal{L}Q)$ and $\partial_zQ(t,z^{\star}(t))=0$, we have 
\begin{eqnarray*}
	\begin{split}
		\begin{cases}
			&-\partial_t (z\partial_z Q) -\mathcal{L} (z\partial_zQ) =\dfrac{1}{\gamma}z^{\frac{1}{\gamma}-1},\;\;\;\mbox{in}\;\Omega_2,\\
			&(z\partial_z Q)(T,z)=0,\;\;\forall\;z\ge z^{\star}(T);\qquad (z\partial_zQ)(t,z^{\star}(t))=0,\;\;\forall\;t\in[0,T].
		\end{cases}
	\end{split}
\end{eqnarray*}
Let us temporarily denote 
$$
Q_1(t,z)=\dfrac{1}{\gamma}\cdot\dfrac{1-e^{-K(T-t)}}{K}z^{\frac{1}{\gamma}-1}.
$$
Then, $Q_1(t,z)$ satisfies 
\begin{eqnarray*}
	\begin{split}
		\begin{cases}
			&-\partial_t Q_1 -\mathcal{L} Q_1 =\dfrac{1}{\gamma}z^{\frac{1}{\gamma}-1},\;\;\;\mbox{in}\;\Omega_2,\\
			&Q_1(T,z)=0,\;\;\forall\;z\ge z^{\star}(T);\; Q_1(t,z^{\star}(t))=\dfrac{1}{\gamma}\cdot\dfrac{1-e^{-K(T-t)}}{K}(z^{\star}(t))^{\frac{1}{\gamma}-1},\;\;\forall\;t\in[0,T].
		\end{cases}
	\end{split}
\end{eqnarray*}
By the comparison principle for PDEs(see \citet{Lieberman}),
$$
z\partial_z Q(t,z)\le Q_1(t,z).
$$
From Lemma \ref{thm:lemQ}, we can conclude
$$
0\le \partial_zQ(t,z) \le \dfrac{1}{\gamma}\cdot\dfrac{1-e^{-K(T-t)}}{K}z^{\frac{1}{\gamma}-2}.
$$
\hfill$\Box$\medskip

{We provide the integral equation representation of $Q(t,z)$ in the following lemma.}
\begin{lem}\label{lem:integral_Q}~
	In the region $\Omega_2$, the value function $Q(t,z)$ has the following integral equation representation:
	\begin{eqnarray*}
		\begin{split}
			Q(t,z)=&\dfrac{z^{\frac{1}{\gamma}-1}}{1-\gamma}\int_{t}^{T}e^{-K(s-t)}\mathcal{N}\left(d^{\gamma}\left(s-t,\frac{z}{z^{\star}(s)}\right)\right)ds\\-&\dfrac{1}{1-\gamma}\int_{t}^{T}e^{-\hat{\rho}(s-t)}\mathcal{N}\left(d^1\left(s-t,\frac{z}{z^{\star}(s)}\right)\right)ds,
		\end{split}
	\end{eqnarray*}
	where 
	$$
	d^1(t,z)=\dfrac{\log{z}+(\hat{r}-\hat{\rho}+\frac{1}{2}(\gamma\sigma)^2)t}{\gamma\sigma\sqrt{t}},\;\;d^\gamma(t,z)=\dfrac{\log{z}+(\hat{r}-\hat{\rho}-\frac{1}{2}(\gamma\sigma)^2+\frac{1}{\gamma}(\gamma\sigma)^2)t}{\gamma\sigma\sqrt{t}},
	$$
	and $\mathcal{N}(\cdot)$ is a standard normal distribution function.
	
	Moreover, the free boundary $z^{\star}(t)$ satisfies the following integral equation:
	\begin{eqnarray*}
		\begin{split}
			0=&\dfrac{(z^{\star}(t))^{\frac{1}{\gamma}-1}}{1-\gamma}\int_{t}^{T}e^{-K(s-t)}\mathcal{N}\left(d^{\gamma}\left(s-t,\frac{z^{\star}(t)}{z^{\star}(s)}\right)\right)ds\\-&\dfrac{1}{1-\gamma}\int_{t}^{T}e^{-\hat{\rho}(s-t)}\mathcal{N}\left(d^1\left(s-t,\frac{z^{\star}(t)}{z^{\star}(s)}\right)\right)ds.
		\end{split}
	\end{eqnarray*}
\end{lem}
\noindent{\bf Proof.} 
From Lemma \ref{thm:lemQ}, 
$$
Q \in W^{1,2}_{p,loc}(\mathcal{M}_T) \cap C(\widetilde{\mathcal{M_T}}),~~~ {p \geq 1 }.
$$
By applying It\'o lemma to $e^{-\hat{\rho} s}Q(s.\mathcal{H}_s)$ (see \citet{Krylov}),
\begin{eqnarray}\begin{split}\label{ito_Q}
\int_{t}^{T}d\left(e^{-{\hat{\rho}}s}Q(s,\mathcal{H}_s)\right)=\int_{t}^{T}e^{-{\hat{\rho}}s}\left(\dfrac{\partial Q}{\partial s} + \mathcal{L}Q\right)ds - {\gamma \sigma}\int_{t}^{T} e^{-{\hat{\rho}}s}\mathcal{H}_s \dfrac{\partial Q}{\partial z}dB^{\mathbb{Q}}_{s}.
\end{split}\end{eqnarray}
By Lemma \ref{thm:estimate},
\begin{eqnarray*}
	\begin{split}
		\mathbb{E}^{\mathbb{Q}}\left[\int_{0}^{T}\left(\gamma\sigma e^{-\hat{\rho}t}\mathcal{H}_t\dfrac{\partial Q}{\partial z}\right)^2dt\right]\le \dfrac{\sigma^2}{K^2} \mathbb{E}^{\mathbb{Q}}\left[\int_{0}^{T}\left(e^{-\hat{\rho}t}{\mathcal{H}_t}^{\frac{\gamma-1}{\gamma}}\right)^2 dt\right] <\infty.
	\end{split}
\end{eqnarray*}
This implies that 
$$
{\gamma \sigma}\int_{t}^{T} e^{-{\hat{\rho}}s}\mathcal{H}_s \dfrac{\partial Q}{\partial z}dB^{\mathbb{Q}}_{s}
$$
is a martingale under $\mathbb{Q}$ measure and 
$$
\mathbb{E}_t^{\mathbb{Q}}\left[ {\gamma \sigma}\int_{t}^{T} e^{-{\hat{\rho}}s}\mathcal{H}_s \dfrac{\partial Q}{\partial z}dB^{\mathbb{Q}}_{s}\right]=0.
$$
(see Chapter 3 in \citet{OS})

By taking expectation to both-side of the equation \eqref{ito_Q}, 
\begin{eqnarray*}
	\begin{split}
		Q(t,z)=&\mathbb{E}_t^{\mathbb{Q}}\left[e^{-\hat{\rho}(T-t)}Q(T,\mathcal{H}_T)\right]-\mathbb{E}_t^{\mathbb{Q}}\left[\int_{t}^{T}e^{-{\hat{\rho}}(s-t)}\left(\dfrac{\partial Q}{\partial s} + \mathcal{L}Q\right)ds\right]\\
		=&-\mathbb{E}^{\mathbb{Q}}_t\left[\int_{t}^{T}e^{-{\hat{\rho}}(s-t)}\left(\dfrac{\partial Q}{\partial s} + \mathcal{L}Q\right){\bf 1}_{\{(s,\mathcal{H}_s)\in{\Omega_2}\}}ds\right]\\
		=&-\mathbb{E}^{\mathbb{Q}}_t\left[\int_{t}^{T}e^{-{\hat{\rho}}(s-t)}\left(\dfrac{\partial Q}{\partial s} + \mathcal{L}Q\right){\bf 1}_{\{\mathcal{H}_s\ge z^{\star}(s)\}}ds\right]\\
		=&\dfrac{1}{1-\gamma}\mathbb{E}_t^{\mathbb{Q}}\left[\int_{t}^{T}e^{-{\hat{\rho}}(s-t)}\left({\mathcal{H}_s}^{\frac{1-\gamma}{\gamma}}-1\right){\bf 1}_{\{\mathcal{H}_s\ge z^{\star}(s)\}}ds\right]\\
		=&\dfrac{1}{1-\gamma}\mathbb{E}_t^{\mathbb{Q}}\left[\int_{t}^{T}e^{-{\hat{\rho}}(s-t)}{\mathcal{H}_s}^{\frac{1-\gamma}{\gamma}}{\bf 1}_{\{\mathcal{H}_s\ge z^{\star}(s)\}}ds\right]-\dfrac{1}{1-\gamma}\mathbb{E}^{\mathbb{Q}}_t\left[\int_{t}^{T}e^{-{\hat{\rho}}(s-t)}{\bf 1}_{\{\mathcal{H}_s\ge z^{\star}(s)\}}ds\right].
	\end{split}
\end{eqnarray*}
Since 
$$
\dfrac{d\mathbb{P}}{d\mathbb{Q}}=\exp{\left\{-\dfrac{1}{2}(1-\gamma)^2\sigma^2(s-t)-(1-\gamma)\sigma(B_s^{\mathbb{Q}}-B_t^{\mathbb{Q}})\right\}}
$$
and $B_s=B_t^{\mathbb{Q}}+(1-\gamma)\sigma s,\;\;\mbox{for}\;s\in[t,T]$,
\begin{eqnarray*}
	\begin{split}
		&\mathbb{E}^{\mathbb{Q}}_t\left[\int_{t}^{T}e^{-{\hat{\rho}}(s-t)}{\mathcal{H}_s}^{\frac{1-\gamma}{\gamma}}{\bf 1}_{\{\mathcal{H}_s\ge z^{\star}(s)\}}ds\right]\\=&z^{\frac{1}{\gamma}-1}\mathbb{E}_t\left[\int_{t}^{T}e^{-K(s-t)}{\bf 1}_{\{\mathcal{H}_s\ge z^{\star}(s)\}}ds\right]\\
		=&z^{\frac{1}{\gamma}-1}\int_{t}^{T}e^{-K(s-t)}\mathbb{P}(\mathcal{H}_s\ge z^{\star}(s))ds\\
		=&z^{\frac{1}{\gamma}-1}\int_{t}^{T}e^{-K(s-t)}\mathcal{N}\left(\dfrac{\log{\frac{z}{z^{\star}(s)}}+(\hat{r}-\hat{\rho}-\frac{1}{2}(\gamma\sigma)^2+\frac{1}{\gamma}(\gamma\sigma)^2)(s-t)}{\gamma\sigma\sqrt{s-t}}\right)ds.
	\end{split}
\end{eqnarray*}
Similarly,
\begin{eqnarray*}
	\begin{split}
		&\mathbb{E}_t^{\mathbb{Q}}\left[\int_{t}^{T}e^{-{\hat{\rho}}(s-t)}{\bf 1}_{\{\mathcal{H}_s\ge z^{\star}(s)\}}ds\right]\\=&\int_{t}^{T}e^{-\hat{\rho}(s-t)}\mathbb{Q}(\mathcal{H}_s\ge z^{\star}(s))ds\\
		=&\int_{t}^{T}e^{-\hat{\rho}(s-t)}\mathcal{N}\left(\dfrac{\log{\frac{z}{z^{\star}(s)}}+(\hat{r}-\hat{\rho}+\frac{1}{2}(\gamma\sigma)^2)(s-t)}{\gamma\sigma\sqrt{s-t}}\right)ds.
	\end{split}
\end{eqnarray*}
Thus, 
\begin{eqnarray*}
	\begin{split}
		Q(t,z)=&\dfrac{z^{\frac{1}{\gamma}-1}}{1-\gamma}\int_{t}^{T}e^{-K(s-t)}\mathcal{N}\left(\dfrac{\log{\frac{z}{z^{\star}(s)}}+(\hat{r}-\hat{\rho}-\frac{1}{2}(\gamma\sigma)^2+\frac{1}{\gamma}(\gamma\sigma)^2)(s-t)}{\gamma\sigma\sqrt{s-t}}\right)ds\\
		-&\frac{1}{1-\gamma}\int_{t}^{T}e^{-\hat{\rho}(s-t)}\mathcal{N}\left(\dfrac{\log{\frac{z}{z^{\star}(s)}}+(\hat{r}-\hat{\rho}+\frac{1}{2}(\gamma\sigma)^2)(s-t)}{\gamma\sigma\sqrt{s-t}}\right)ds.
	\end{split}
\end{eqnarray*}
By the smooth-pasting condition (Lemma \ref{lem-cinfinity}),
\begin{eqnarray*}
	\begin{split}
		0=&(z^{\star}(t))^{\frac{1}{\gamma}-1}\int_{t}^{T}e^{-K(s-t)}\mathcal{N}\left(\dfrac{\log{\frac{z^{\star}(t)}{z^{\star}(s)}}+(\hat{r}-\hat{\rho}-\frac{1}{2}(\gamma\sigma)^2+\frac{1}{\gamma}(\gamma\sigma)^2)(s-t)}{\gamma\sigma\sqrt{s-t}}\right)ds\\
		-&\int_{t}^{T}e^{-\hat{\rho}(s-t)}\mathcal{N}\left(\dfrac{\log{\frac{z^{\star}(t)}{z^{\star}(s)}}+(\hat{r}-\hat{\rho}+\frac{1}{2}(\gamma\sigma)^2)(s-t)}{\gamma\sigma\sqrt{s-t}}\right)ds.
	\end{split}
\end{eqnarray*}
\hfill$\Box$\medskip

{We will show that the value function $Q(t,z)$ converges when the time-to-maturity goes to infinity. We will need the following lemma.  }
\begin{lem} ~For arbitrary $c>0$ and $d \in\mathbb{R}$, \label{lem:lemcd}
	\begin{eqnarray*}
		\int_0^{\infty}e^{-c\xi}\mathcal{N}(d\sqrt{\xi})d\xi = \frac{1}{2c}\left(1+\frac{d}{\sqrt{d^2+2c}}\right).
	\end{eqnarray*}
\end{lem}
\noindent{\bf Proof.}  By integration by parts,
\begin{eqnarray}
\int_0^{\infty}e^{-c\xi}\mathcal{N}(d\sqrt{\xi})d\xi =\left[-\frac{1}{c} e^{-c \xi}\mathcal{N}(d\sqrt{\xi})\right]_{\xi=0}^{\xi=\infty} + \frac{d}{2c \sqrt{2\pi}}\int_0^{\infty} e^{-c\xi -\frac{d^2}{2}\xi}\frac{1}{\sqrt{\xi}}d\xi.
\end{eqnarray}
By \citet{AS} (p.304, equation (7.4.33)), for any $a,b \in \mathbb{R}$,
\begin{eqnarray*}
	\int_0^{\infty}\exp\left\{ -a^2x^2-\frac{b^2}{x^2} \right\}dx = \frac{\sqrt{\pi}}{2|a|}e^{-2|a||b|}.
\end{eqnarray*}
Therefore,
\begin{eqnarray*}
	\int_0^{\infty} e^{-(c +\frac{d^2}{2})\xi}\frac{1}{\sqrt{\xi}}d\xi=\frac{\sqrt{\pi}}{\sqrt{c+\frac{d^2}{2}}}.
\end{eqnarray*}
\hfill$\Box$\medskip

{We now provide the convergence of $Q(t,z)$ in the following lemma.}
\begin{lem}\label{lem:inf}
	\begin{eqnarray*}
		\begin{split}
			\lim_{T-t \to \infty} Q(t,z)=Q_{\infty}(z)\;\;\;\mbox{and}\;\;\lim_{T-t\to \infty}z^{\star}(t)=z_\infty,
		\end{split}
	\end{eqnarray*}
	where $Q_{\infty}(z)$ and $z_\infty$ are defined in \eqref{eq:INFQ} and \eqref{eq:INFZ}, respectively.
\end{lem}
\noindent{\bf Proof.} 
\noindent Let $\widetilde{z}_{\star}$ define such that $\widetilde{z}_{\star}(T-t) \equiv z^{\star}(t)$, then 
\begin{eqnarray*}
	z_\infty = \lim_{T-t \rightarrow \infty} z^{\star}(t)=\lim_{T-t \rightarrow \infty} \widetilde{z}_{\star}(T-t).
\end{eqnarray*}
\noindent From the integral equation of $z^{\star}$ in Lemma \ref{lem:integral_Q} and $T-t \rightarrow \infty$,
\begin{eqnarray} \label{eq:yinfty}
\begin{split}
0&= {z_{\infty}}^{\frac{\gamma-1}{\gamma}} \int_0^\infty e^{-K\xi} \mathcal{N}\left(\frac{\hat{r}-\hat{\rho}-\frac{1}{2}(\gamma \sigma)^2+\frac{1}{\gamma}(\gamma \sigma)^2 }{\gamma \sigma}\sqrt{\xi}\right) d\xi\\
&~-\int_0^\infty e^{-\hat{\rho}\xi} \mathcal{N}\left(\frac{\hat{r}-\hat{\rho}+\frac{1}{2}(\gamma \sigma)^2 }{\gamma \sigma}\sqrt{\xi}\right) d\xi.\\
\end{split}
\end{eqnarray}
By applying Lemma \ref{lem:lemcd} and direct computation, we can get $z_\infty$ as follows:
\begin{eqnarray*}
	z_{\infty}=\left(\frac{\hat{\rho}(\alpha_- \gamma + \gamma -1)}{K \alpha_- \gamma}\right)^{\frac{\gamma}{\gamma-1}}.
\end{eqnarray*}
Consider the integral equation representation of $Q(t,y)$ in Lemma \ref{lem:integral_Q}, then
\begin{eqnarray*}
	\begin{aligned}
		Q_{\infty}(z) \equiv &\lim_{T-t \rightarrow \infty}Q(t,z)\\
		=&\lim_{T-t \rightarrow \infty}\dfrac{z^{\frac{1}{\gamma}-1}}{1-\gamma}\int_{0}^{T-t}e^{-K\xi}\mathcal{N}\left(d^{\gamma}\left(\xi,\frac{z}{\widetilde{z}^{\star}(T-t-\xi)}\right)\right)d\xi\\
		&-\lim_{T-t \rightarrow \infty} \dfrac{1}{1-\gamma}\int_{0}^{T-t}e^{-\hat{\rho}\xi}\mathcal{N}\left(d^1\left(\xi,\frac{z}{\widetilde{z}^{\star}(T-t-\xi)}\right)\right)d\xi\\
		=& \dfrac{z^{\frac{1}{\gamma}-1}}{1-\gamma}\int_{0}^{\infty}e^{-K\xi}\mathcal{N}\left(d^{\gamma}\left(\xi,\frac{z}{z_\infty}\right)\right)d\xi - \dfrac{1}{1-\gamma}\int_{0}^{\infty}e^{-\hat{\rho}\xi}\mathcal{N}\left(d^1\left(\xi,\frac{z}{z_\infty}\right)\right)d\xi \\
		=&\dfrac{z^{\frac{1}{\gamma}-1}}{1-\gamma}\int_{0}^{\infty}e^{-K\xi}\mathcal{N}\left(\left(\dfrac{\log\frac{z}{z_\infty}}{\gamma \sigma}\right) \xi^{-\frac{1}{2}}+\left(\dfrac{\hat{r}-\hat{\rho}-\frac{1}{2}(\gamma \sigma)^2+\frac{1}{\gamma}(\gamma \sigma)^2}{\gamma \sigma}\right)\xi^{\frac{1}{2}}\right)d\xi \\
		&-\dfrac{1}{1-\gamma}\int_{0}^{\infty}e^{-\hat{\rho}\xi}\mathcal{N}\left(\left(\dfrac{\log\frac{z}{z_\infty}}{\gamma \sigma}\right)\xi^{-\frac{1}{2}}+\left(\dfrac{\hat{r}-\hat{\rho}+\frac{1}{2}(\gamma \sigma)^2}{\gamma \sigma}\right)\xi^{\frac{1}{2}} \right)d\xi.
	\end{aligned}
\end{eqnarray*}
The exact solution of the above integral equation can be easily calculated by integration by parts and some characteristic solutions. 

First, consider the following integral equation: 
\begin{eqnarray*}
	\int_{0}^{\infty}e^{-K\xi}\mathcal{N}\left(\left(\dfrac{\log\frac{z}{z_\infty}}{\gamma \sigma}\right) \xi^{-\frac{1}{2}}+\left(\dfrac{\hat{r}-\hat{\rho}-\frac{1}{2}(\gamma \sigma)^2+\frac{1}{\gamma}(\gamma \sigma)^2}{\gamma \sigma}\right)\xi^{\frac{1}{2}}\right)d\xi. 
\end{eqnarray*}	
Let
\begin{eqnarray}
c_1=\left(\dfrac{\log\frac{z}{z_\infty}}{\gamma \sigma}\right),\;\;\mbox{and}\;\;c_2=\left(\dfrac{\hat{r}-\hat{\rho}+\frac{1}{2}(\gamma \sigma)^2}{\gamma \sigma}\right).
\end{eqnarray}
Since we consider the value of $Q$ in the $\Omega_2$,  $z\geq z^\star(\xi+t) \geq z_\infty$ implies  $c_1>0$.\\ 
By integration by parts,
\begin{eqnarray*}
	\begin{aligned}
		&\int_0^\infty e^{-K \xi} \mathcal{N}\left(c_1 \xi^{-\frac{1}{2}}+c_2 \xi^{\frac{1}{2}} \right)\\
		=& \left.-\frac{1}{K}e^{-K\xi}\mathcal{N}\left(c_1 \xi^{-\frac{1}{2}}+c_2\xi^{\frac{1}{2}}\right)\right]_{\xi=0}^\infty\\
		&+\frac{1}{K}\frac{1}{\sqrt{2\pi}}\int_0^\infty e^{-K\xi}\cdot e^{\frac{1}{2}(c_1 \xi^{-\frac{1}{2}}+c_2 \xi^{\frac{1}{2}})^2}\cdot \left(-\frac{c_1}{2}\xi^{-\frac{3}{2}}+\frac{c_2}{2}\xi^{-\frac{1}{2}}\right)d\xi\\
		=&\frac{1}{K}+\frac{e^{-c_1c_2}}{K \sqrt{2\pi}}\int_0^\infty e^{-\left(
			\frac{c_1}{\sqrt{2}}\right)^2x^{-2} -\left(\frac{\sqrt{c_2^2+2K}}{\sqrt{2}}\right)^2x^2}(-c_1x^{-2}+c_2)dx.~~~~(\xi=x^2)\\
	\end{aligned}
\end{eqnarray*}
Since \citet{AS} (p.304, equation (7.4.33)),
\begin{eqnarray*}
	\begin{split}
		&\int_0^{\infty} \exp \left\{-a^2x^2-\frac{b^2}{x^2}\right\}dx = \frac{\sqrt{\pi}}{2|a|}e^{-2|a||b|},\\
		&\int_0^{\infty} \frac{1}{x^2} \exp \left\{-a^2x^2-\frac{b^2}{x^2}\right\}dx = \frac{\sqrt{\pi}}{2|b|}e^{-2|a||b|}.\\
	\end{split}
\end{eqnarray*}
By using above equations , we can get
\begin{eqnarray}\label{eq:INF1}
\begin{aligned}
\int_0^\infty e^{-K \xi} \mathcal{N}\left(c_1 \xi^{-\frac{1}{2}}+c_2 \xi^{\frac{1}{2}} \right)=&\frac{1}{K}-\frac{1}{2K}\left(1-\frac{c_2}{\sqrt{c_2^2 + 2K}}\right)e^{-c_1(\sqrt{c_2^2 + 2K}+c_2)}\\
=&\frac{1}{K}\left(1-\left(\frac{\alpha_++\left(\tfrac{\gamma-1}{\gamma}\right)}{\alpha_+-\alpha_-}\right)\cdot \left(\frac{z}{z_\infty}\right)^{\alpha_-+\left(\tfrac{\gamma-1}{\gamma}\right)}\right),
\end{aligned}
\end{eqnarray}
where
\begin{eqnarray*}
	\begin{split}
		&\sqrt{c_2^2+ 2K} = \frac{\gamma \sigma(\alpha_+-\alpha_-)}{2},\;\;\sqrt{c_2^2+ 2K}+c_2 = -\gamma \sigma(\alpha_-+\left(\tfrac{\gamma-1}{\gamma}\right)),\\
		&\sqrt{c_2^2+ 2K}-c_2 = \gamma \sigma (\alpha_++\left(\tfrac{\gamma-1}{\gamma}\right)).
	\end{split}
\end{eqnarray*}
Similarly, we can derive
\begin{eqnarray}\label{eq:INF2}
\begin{aligned}
&\int_{0}^{\infty}e^{-\hat{\rho}\xi}\mathcal{N}\left(\left(\dfrac{\log\frac{z}{z_\infty}}{\gamma \sigma}\right)\xi^{-\frac{1}{2}}+\left(\dfrac{\hat{r}-\hat{\rho}+\frac{1}{2}(\gamma \sigma)^2}{\gamma \sigma}\right)\xi^{\frac{1}{2}} \right)d\xi
=\frac{1}{\hat{\rho}}\left(1-\left(\frac{\alpha_+}{\alpha_+-\alpha_-}\right)\left(\frac{z}{z_\infty}\right)^{\alpha_-}\right).
\end{aligned}
\end{eqnarray}
By the equation (\ref{eq:INF1}) and (\ref{eq:INF2}), we can get $Q_{\infty}$ as follows:
\begin{eqnarray*}
	\begin{aligned}
		Q_{\infty}(z)=\left(-\dfrac{1}{\gamma}\dfrac{1}{K}\dfrac{1}{\alpha_-}(z_{\infty})^{\frac{1}{\gamma}-1-\alpha_-}\right)z^{\alpha_-}+\dfrac{1}{1-\gamma}\left(\dfrac{1}{K}z^{\frac{1}{\gamma}-1}-\dfrac{1}{\hat{\rho}}\right),~~~~~
		\mbox{in}~~~\Omega_2^{\infty}.
	\end{aligned}
\end{eqnarray*}
\hfill$\Box$\medskip

\section{Duality Theorem and Optimal Strategies}\label{sec:duality}
\begin{figure}[h]
	\centering
	\subfigure{\includegraphics[scale=0.55]{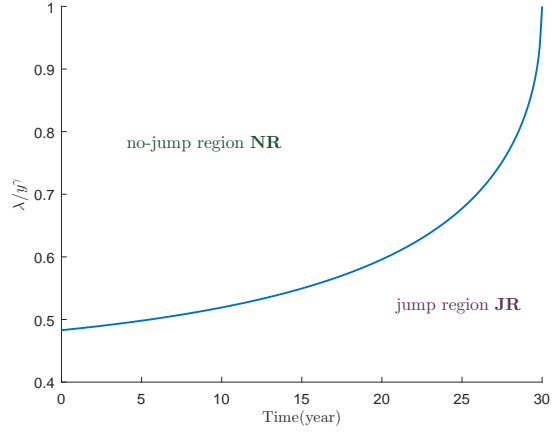}}
	\caption{\label{fig:1} The jump-region and the no-jump region in $(t,z)$-domain.}
\end{figure}

By analyzing the variational inequality \eqref{eq:VI1} in Section \ref{sec:B}, we derive the following integral equation representation of value function of $g(t,z)$ of Problem \ref{pr:OS}:
\begin{eqnarray*}
	\begin{split}
		g(t,z)=&\dfrac{1}{1-\gamma}\int_{t}^{T}e^{-\hat{\rho}(s-t)}\mathcal{N}\left(-d^1\left(s-t,\frac{z}{z^{\star}(s)}\right)\right)ds\\-&\dfrac{z^{\frac{1}{\gamma}-1}}{1-\gamma}\int_{t}^{T}e^{-K(s-t)}\mathcal{N}\left(-d^{\gamma}\left(s-t,\frac{z}{z^{\star}(s)}\right)\right)ds,
	\end{split}
\end{eqnarray*}
where $z^{\star}(t)$ is the free boundary of Problem \ref{pr:OS}, $\mathcal{N}(\cdot)$ is a standard normal distribution function, and
$$
d^1(t,z)=\dfrac{\log{z}+(\hat{r}-\hat{\rho}+\frac{1}{2}(\gamma\sigma)^2)t}{\gamma\sigma\sqrt{t}},\;\;d^\gamma(t,z)=\dfrac{\log{z}+(\hat{r}-\hat{\rho}-\frac{1}{2}(\gamma\sigma)^2+\frac{1}{\gamma}(\gamma\sigma)^2)t}{\gamma\sigma\sqrt{t}}.
$$
Also, in terms of free boundary $z^{\star}(t)$, we can define the jump region {\bf JR} and the no-jump region {\bf NR} as follows:
$$
{\bf JR}=\{(t,\lambda,y) \mid \lambda \le z^{\star}(t) y^\gamma\}\;\;\;\mbox{and}\;\;\;{\bf NR}=\{(t,\lambda,y)\mid \lambda > z^{\star}(t)y^\gamma \}.
$$
As seen in Figure \ref{fig:1}, the free boundary $z^{\star}(t)$ partitions the $(t,z)$-region into the jump region and no-jump region.
\begin{rem}~	In terms of te regions $\Omega_1$ and $\Omega_2$ defined in Section \ref{sec:B},
	$$
	{\bf JR}=\{(t,\lambda,y)\mid (t,\dfrac{\lambda}{y^\gamma})\in \Omega_1\}\;\;\;\textrm{and}\;\;\;{\bf NR}=\{(t,\lambda,y)\mid (t,\dfrac{\lambda}{y^\gamma})\in \Omega_2\}.
	$$
\end{rem}

If initially $(t,\lambda,y)\in {\bf JR}${,} then $X$ should  {increase} immediately, such that $\dfrac{\lambda}{y^\gamma}$ reaches the free boundary $z^{\star}(t)$. That is, the principal should increase the agent's consumption process. On the other hand, if $(t, \lambda,y)\in {\bf NR}$, $X$ must stay constant and this implies that the principal does not adjust the agent's consumption process. Thus, we call {\bf JR}  and {\bf NR} the jump region and the no-jump region, respectively.

Moreover, the free boundary $z^{\star}(t)$ satisfies the following integral equation:
\begin{eqnarray}
\begin{split}\label{eq:integral}
0=&\dfrac{(z^{\star}(t))^{\frac{1}{\gamma}-1}}{1-\gamma}\int_{t}^{T}e^{-K(s-t)}\mathcal{N}\left(d^{\gamma}\left(s-t,\frac{z^{\star}(t)}{z^{\star}(s)}\right)\right)ds\\-&\dfrac{1}{1-\gamma}\int_{t}^{T}e^{-\hat{\rho}(s-t)}\mathcal{N}\left(-d^1\left(s-t,\frac{z^{\star}(t)}{z^{\star}(s)}\right)\right)ds.
\end{split}
\end{eqnarray}
Then, we can directly obtain the following lemma.
\begin{lem}~\label{lem:time}
	The optimal stopping time $\tau^*$ for Problem \ref{pr:OS} is given by 
	$$
	\tau^* \equiv \inf \left\{s \ge t \bigg| \mathcal{H}_s \le z^{\star}(t)  \right\}\wedge T,
	$$
	where $z^{\star}(t)$ satisfies the integral equation \eqref{eq:integral}.
\end{lem}
By using Lemma \ref{thm:lem1} and Lemma \ref{lem:time}, we provide a solution to Problem \ref{pr:dual} in the following proposition.
\begin{pro}~\label{pro:dual}
	\begin{itemize}
		
		\item[(a)] The infinite series of optimal stopping times $\{\tau^*(x)\}_{x\ge \lambda}$ in Lemma \ref{thm:lem1} is given by
		$$
		\tau^*(x) = \inf \{s\ge t \mid  x\mathcal{H}_s \le z^{\star}(t)\} \wedge T.
		$$
		\item[(b)] 	The dual value function is given by 
		\begin{eqnarray*}
			\begin{split}
				J(t,\lambda,y)=&-y^{1-\gamma}\int_{\lambda}^{\infty}g(t,\dfrac{x}{y^{\gamma}})dx + J_0(t,\lambda,y)\\
				=&-\dfrac{y^{1-\gamma}}{1-\gamma}\int_{\lambda}^{\infty}\left[\int_{t}^{T}e^{-\hat{\rho}(s-t)}\mathcal{N}\left(-d^\gamma(s-t,\frac{x}{ z^{\star}(s)y^\gamma})\right)ds\right.\\&\left.-\left(\dfrac{x}{y^\gamma}\right)^{\frac{1}{\gamma}-1}\int_{t}^{T}e^{-\hat{\rho}(s-t)}\mathcal{N}\left(-d^1(s-t,\frac{x}{z^{\star}(s)y^\gamma})\right)ds\right]dx\\
				&+\dfrac{\gamma}{1-\gamma}\dfrac{1-e^{-K(T-t)}}{K}\lambda^{\frac{1}{\gamma}}+\dfrac{1-e^{-\hat{r}(T-t)}}{\hat{r}}y.
			\end{split}
		\end{eqnarray*}
		
	\end{itemize}
\end{pro}

\begin{rem}~ It is easy to see that the infinite series of optimal stopping times $\{\tau^*(x)\}_{x\ge \lambda}$ defined in Proposition  \ref{pro:dual} is non-decreasing, left continuous with right limits as function of $x$. Thus, the optimal stopping problem in \eqref{eq:dual_ex1} can be expressed as follows:
	\begin{eqnarray}
	\sup_{\tau^*(x)\in[t,T]}\mathbb{E}_t^{\mathbb{Q}} \left[ e^{-\hat{\rho}(\tau^*(x)-t)} h(\tau^*(x),x\mathcal{H}_{\tau^*(x)}) \right]=g(t,xH_t).
	\end{eqnarray}
\end{rem}
Proposition \ref{pro:dual} (a) characterizes the optimal time to adjust the process $X$ as we discussed earlier.
Proposition \ref{pro:dual} (b) provides the dual value function by using the time-varying function $z^{\star}(t)$ determining the free boundary for the optimal stopping problem. 

By Proposition \ref{pro:dual}, 
\begin{eqnarray}\label{eq:partial_J}
\partial_\lambda J(t,\lambda,y)&=y^{1-\gamma}g\left(t,\dfrac{\lambda}{y^{\gamma}}\right)+\partial_\lambda J_0(t,\lambda,y).
\end{eqnarray}
From \eqref{eq:def_region1} and \eqref{eq:def_region2} in Section \ref{sec:B},
\begin{eqnarray*}
	\begin{split}
		\partial_\lambda J(t,\lambda,y)&> y^{1-\gamma}h\left(t,\dfrac{\lambda}{y^\gamma}\right)+\partial_\lambda J_0\left(t,\lambda,y\right)=U_d(t,y),\;\;\;\mbox{for}\;\;\lambda>y^{\gamma}z^{\star}(t),\\
		\partial_\lambda J(t,\lambda,y)&= y^{1-\gamma}h\left(t,\dfrac{\lambda}{y^\gamma}\right)+\partial_\lambda J_0(t,\lambda,y)=U_d(t,y),\;\;\;\mbox{for}\;\;\lambda\le y^{\gamma}z^{\star}(t).\\
	\end{split}
\end{eqnarray*}
Hence, we can rewrite 
\begin{eqnarray*}
	\begin{split}
		{\bf JR}&=\{(t,\lambda,y) \mid \partial_{\lambda}J(t,\lambda,y)=U_d(t,y) \},\\
		{\bf NR}&=\{(t,\lambda,y) \mid \partial_{\lambda}J(t,\lambda,y)>U_d(t,y) \}.
	\end{split}
\end{eqnarray*}
Now we can state the following corollary. 
\begin{cor}
	~The dual value function $J(t,\lambda,y)$ can be rewritten by 
	\begin{itemize}
		\item[(a)] In the no-jump region {\bf NR},
		\begin{eqnarray*}
			\begin{split}
				J(t,\lambda,y)=&-\dfrac{y^{1-\gamma}}{1-\gamma}\int_{\lambda}^{\infty}\left[\int_{t}^{T}e^{-\hat{\rho}(s-t)}\mathcal{N}\left(-d^1\left(s-t,\frac{x}{ z^{\star}(s)y^\gamma}\right)\right)ds\right.\\&\left.-\left(\dfrac{x}{y^\gamma}\right)^{\frac{1}{\gamma}-1}\int_{t}^{T}e^{-\hat{\rho}(s-t)}\mathcal{N}\left(-d^\gamma \left(s-t,\frac{x}{z^{\star}(s)y^\gamma}\right)\right)ds\right]dx\\
				&+\dfrac{\gamma}{1-\gamma}\cdot \dfrac{1-e^{-K(T-t)}}{K}\lambda^{\frac{1}{\gamma}}+\dfrac{1-e^{-\hat{r}(T-t)}}{\hat{r}}y.
			\end{split}
		\end{eqnarray*}
		\item[(b)]  {In} the jump-region {\bf JR},
		\begin{eqnarray*}
			\begin{split}
				J(t,\lambda,y)=J(t,z^{\star}(t)y^{\gamma},y) + (\lambda-z^{\star}(t)y^{\gamma})U_d(t,y).
			\end{split}
		\end{eqnarray*}
	\end{itemize}
\end{cor}
By applying a standard method of singular control problem developed by
\citet{DN} or \citet{FS} to Problem \ref{pr:dual}, the dual value function $J(t,\lambda,y)$ satisfies the certain Hamilton-Jacobi-Bellman(HJB) equation. In fact, \citet{MJ} study the infinite-horizon problem by solving the linear Hamilton-Jacobi-Bellman equation. The following proposition provides that our derived dual value function $J(t,\lambda,y)$ satisfies the associated HJB equation.
\begin{pro}\label{pro:dual_HJB}~
	The dual value function $J(t,\lambda,y)$ satisfies the following HJB equation: 
	\begin{eqnarray*}
		\begin{split}
			\begin{cases}
				&\min\left\{\partial_t J+ \mathcal{A}J+y+\tilde{u}(\lambda),\;\partial_\lambda J-U_d(t,y)\right\}=0,\;\;\;(t,\lambda,y)\in[0,T]\times\mathbb{R}_{+}\times\mathbb{R}_+,\\
				&J(T,\lambda,y)=0,
			\end{cases}
		\end{split}
	\end{eqnarray*}
	where 
	$$
	\mathcal{A}=\frac{\sigma^2}{2}y^2 \partial_{yy}+\mu y\partial_y+(r-\rho)\lambda\partial_{\lambda} -r.
	$$
\end{pro}
\noindent{\bf Proof.} 
It is sufficient to 
$$
\partial_t J+ \mathcal{A}J+y+\tilde{u}(\lambda)=0 \;\;\;\mbox{in}\;\; {\bf NR},
$$ 
and 
$$
\partial_t J+\mathcal{A}J+y+\tilde{u}(\lambda)\ge 0 \;\;\;\mbox{in}\;\; {\bf JR}.
$$
From the representation of $J(t,\lambda,y)$ in Proposition \ref{pro:dual}, we have 
\begin{eqnarray}\label{eq:DER_J}
\begin{split}
\partial_tJ&=y^{1-\gamma}\int_\lambda^\infty -\partial_tg\left(t,\dfrac{x}{y^{\gamma}}\right)dx -\frac{\gamma}{1-\gamma}e^{-K(T-t)}\lambda^{\frac{1}{\gamma}}-e^{\hat{r}(T-t)}y\\
y \partial_y J &= y^{1-\gamma}\int_\lambda^\infty \Big[-(1-\gamma) g\left(t,\dfrac{x}{y^{\gamma}}\right) + \gamma \dfrac{x}{y^{\gamma}}\partial_z g\left(t,\dfrac{x}{y^{\gamma}}\right)\Big]dx\\
y^2 \partial_{yy}J &=y^{1-\gamma}\int_\lambda^\infty \Big[\gamma(1-\gamma)g\left(t,\dfrac{x}{y^{\gamma}}\right)-\left((\gamma^2-\gamma)+ 2\gamma^2 \right)\dfrac{x}{y^{\gamma}}\partial_z g\left(t,\dfrac{x}{y^{\gamma}}\right) \Big.\\
&\Big.-\gamma^2  \left(\dfrac{x}{y^{\gamma}}\right)^2\partial_{zz} g\left(t,\dfrac{x}{y^{\gamma}}\right)\Big]dx\\
\lambda \partial_\lambda J&=y^{1-\gamma}\lambda g\left(t,\frac{\lambda}{y^\gamma}\right)+\frac{1}{1-\gamma}\frac{1-e^{-K(T-t)}}{K}\lambda^{\frac{1}{\gamma}}.
\end{split}
\end{eqnarray}
Since the value function $g(t,z)$ satisfies the variational inequality \eqref{eq:VI1},
\begin{eqnarray}\label{eq:NR_G}
-\partial_t g(t,\dfrac{x}{y^{\gamma}})-\mathcal{L}g(t,\dfrac{x}{y^\gamma})=0,~~~\textrm{for}\;\;(x,y)\in{\bf NR}.
\end{eqnarray}
By using (\ref{eq:DER_J}) and (\ref{eq:NR_G}), $\mathcal{A}J(t,\lambda,y)$ can be derived as follows:
\begin{eqnarray}
\begin{aligned}
&\partial_t J(t,\lambda,y)+\mathcal{A}J(t,\lambda,y)\\
=&y^{1-\gamma}\int_\lambda^\infty \Big[-\partial_t g-\frac{\gamma^2\sigma^2}{2}\left(\dfrac{x}{y^{\gamma}}\right)^2\partial_{zz}g-(\hat{r}-\hat{\rho}+\gamma^2\sigma^2+(\rho-r))\dfrac{x}{y^{\gamma}}\partial_z g+(\hat{\rho}-(\rho-r))g \Big]dx\\
&-y^{1-\gamma}(\rho-r)\lambda g\left(t,\frac{\lambda}{y^\gamma}\right)-\frac{\gamma}{1-\gamma}\lambda^{\frac{1}{\gamma}}-y\\
=&y^{1-\gamma}\int_\lambda^\infty \Big[ -\partial_t g-\mathcal{L}g(t,\dfrac{x}{y^{\gamma}})-(\rho-r)\Big(\dfrac{x}{y^{\gamma}}\partial_z g(t,\dfrac{x}{y^{\gamma}})+g(t,\dfrac{x}{y^{\gamma}}) \Big)\Big]dx\\
&-y^{1-\gamma}(\rho-r)\lambda g\left(t,\frac{\lambda}{y^\gamma}\right)-\frac{\gamma}{1-\gamma}\lambda^{\frac{1}{\gamma}}-y\\
=& -y^{1-\gamma}(\rho-r)\left(\int_\lambda^\infty \left[\frac{x}{y^\gamma}\partial_zg\left(t,\frac{x}{y^\gamma}\right)+g\left(t,\frac{x}{y^{\gamma}}\right)\right]dx+\lambda g\left(t,\frac{\lambda}{y^\gamma}\right)\right)-\frac{\gamma}{1-\gamma}\lambda^{\frac{1}{\gamma}}-y\\
=& -y^{1-\gamma}(\rho-r)\left( \left[xg(t,\dfrac{x}{y^\gamma})\right]_{x=\lambda}^{\infty}+\lambda g\left(t,\frac{\lambda}{y^\gamma}\right)\right)-\frac{\gamma}{1-\gamma}\lambda^{\frac{1}{\gamma}}-y.
\end{aligned}
\end{eqnarray}
By Lemma \ref{lem:integral_Q} and $g(t,z)=Q(t,z)+h(t,z)$, we have 
\begin{eqnarray*}
	\begin{split}
		g(t,z)=&\dfrac{z^{\frac{1}{\gamma}-1}}{1-\gamma}\int_{t}^{T}e^{-K(s-t)}\mathcal{N}\left(-d^{\gamma}\left(s-t,\frac{z}{z^{\star}(s)}\right)\right)ds\\-&\dfrac{1}{1-\gamma}\int_{t}^{T}e^{-\hat{\rho}(s-t)}\mathcal{N}\left(-d^1\left(s-t,\frac{z}{z^{\star}(s)}\right)\right)ds,
	\end{split}
\end{eqnarray*}
Since $\mathcal{N}\left(-d^{\gamma}\left(s-t,\frac{z}{z^{\star}(s)}\right)\right)$ exponentially converge to $0$ {as} $z$ goes to infinity, it is easy to show that 
$$
\lim_{z\to \infty} zg(t,z)=0.
$$
This implies that 
\begin{eqnarray*}
	\begin{split}
		\partial_t J(t,\lambda,y)+\mathcal{A}J(t,\lambda,y)=&-y^{1-\gamma}(\rho-r)\left( \left[xg(t,\dfrac{x}{y^\gamma})\right]_{x=\lambda}^{\infty}+\lambda g\left(t,\frac{\lambda}{y^\gamma}\right)\right)-\frac{\gamma}{1-\gamma}\lambda^{\frac{1}{\gamma}}-y\\
		=&-\frac{\gamma}{1-\gamma}\lambda^{\frac{1}{\gamma}}-y.
	\end{split}
\end{eqnarray*}
and
\begin{eqnarray*}
	\partial_t J(t,\lambda,y)+\mathcal{A} J(t,\lambda,y)+y+\tilde{u}(\lambda)=0.
\end{eqnarray*}
Since 
\begin{eqnarray}\label{eq:JR_G}
-\partial_t g(t,\dfrac{x}{y^{\gamma}})-\mathcal{L}g(t,\dfrac{x}{y^\gamma})\ge 0~~~\textrm{for}\;\;(x,y)\in{\bf JR},
\end{eqnarray}
we can similarly show that 
\begin{eqnarray*}
	\partial_t J(t,\lambda,y)+\mathcal{A} J(t,\lambda,y)+y+\tilde{u}(\lambda)\ge 0\;\;\mbox{in}\;\;{\bf JR}.
\end{eqnarray*}
\hfill$\Box$\medskip

We will now state and prove the main theorem of this paper.
\begin{thm}~\label{thm:main}
	\begin{itemize}
		\item[(a)] For given $w$ satisfying Assumption \ref{as:promise_value}, the value function $V(t,w,y)$ of Problem \ref{pr:main} and the dual value function $J(t,\lambda,y)$ derived in Proposition \ref{pro:dual} satisfy the following duality relationship {:}
		\begin{eqnarray}
		\begin{split}\label{eq:duality}
		V(t,w,y)=\min_{\lambda >0}\left(J(t,\lambda,y)-\lambda w\right).
		\end{split}
		\end{eqnarray}
		There exists a unique solution $\lambda^*$  with $(t,\lambda^*,y)\in {\bf NR}$ for the minimization problem \eqref{eq:duality}.
		\item[(b)] For $s\in[t,T]$, the optimal costate process $X_s^*$ is given by 
		\begin{equation}
		X_s^*=\max\left(\lambda^*, \sup_{t\le \xi \le s}e^{(\rho-r)(\xi-t)}Y_{\xi}^\gamma z^{\star}(\xi)\right),	
		\end{equation}
		where $\lambda^*$ is the unique solution to the minimization problem \eqref{eq:duality} in (a). Moreover, the optimal costate process $\{X_s^*\}_{s=t}^{T}$ satisfies the integrability condition \eqref{eq:integrability}.
		\item[(c)] For $s\in[t,T]$, the optimal consumption plan $c_s^*$ and  continuation value $w_t^*$ are, respectively, given by 
		\begin{eqnarray*}
			\begin{split}
				C_s^*=& \left(\lambda_s^*\right)^{\frac{1}{\gamma}},\\
				w_s^*=& \dfrac{{(\lambda_s^*)}^{1-\gamma}}{1-\gamma}\int_{t}^{T}e^{-\hat{\rho}(s-t)}\mathcal{N}\left(-d^1\left(s-t,\dfrac{\lambda_s^*}{y_s^{\gamma}z^{\star}(t)}\right)\right)ds\\+&\dfrac{(\lambda_s^*)^{\frac{1}{\gamma}-1}}{1-\gamma}\int_{t}^{T}e^{-K(s-t)}\mathcal{N}\left(d^{\gamma}\left(s-t,\dfrac{\lambda_s^*}{y_s^\gamma z^\star(s)}\right)\right)ds
			\end{split}
		\end{eqnarray*}
		with $\lambda_s^*\equiv e^{-(\rho-r)(s-t)}X_s^*$.
	\end{itemize}
\end{thm}
\noindent{\bf Proof.} 
We will prove the duality relationship in the theorem in the following steps.\\

\noindent {\bf (Step 1)} First, we will show that the dual value function $J(t,\lambda,y)$ is strictly convex in $\lambda$ in {\bf NR}.\\

\noindent{\bf Proof of {\bf (Step 1)}}  Consider $\lambda^1,\lambda^2>0\;(\lambda^1\neq \lambda^2)$ with 
$$
(t,\lambda^1,y),\;(t,\lambda^2,y)\in{\bf NR}.
$$
Let $\lambda^3=\alpha \lambda^1 + (1-\alpha)\lambda^2$ with $\alpha\in(0,1)$. Then, clearly 
$$
(t,\lambda^3,y)\in{\bf NR}.
$$
Also, let $X^{*,j}$ be the optimal process for the minimization problem {\eqref{eq:dual_value}} with $\{X_s^{*,j}\}_{s=t}^{T}\in\mathcal{ND}(\lambda^{j})$, $j=1,2,3$. In other words, for $j=1,2,3$,
\begin{eqnarray*}
	\begin{split}
		J(t,\lambda^{j},y)=& \inf_{X \in \mathcal{ND}(\lambda^j)} \left\{\mathbb{E}\left[\int_t^T e^{-r(s-t)} \Big( \tilde{u}(e^{-(\rho-r)(s-t)}X_s) +Y_s - e^{-(\rho-r)(s-t)}X_s u(Y_s)\Big)ds  \right] \right. \\
		+&\left. \lambda^j \mathbb{E}_t \left[\int_t^T e^{-\rho(s-t)}u(Y_s)ds \right] \right\}\\
		=&\mathbb{E}\left[\int_t^T e^{-r(s-t)} \Big( \tilde{u}(e^{-(\rho-r)(s-t)}X_s^{*,j}) +Y_s - e^{-(\rho-r)(s-t)}X_s^{*,j} u(Y_s)\Big)ds  \right]  \\
		+&\lambda^j \mathbb{E}_t \left[\int_t^T e^{-\rho(s-t)}u(Y_s)ds \right].
	\end{split}
\end{eqnarray*}
Then, 
\begin{eqnarray*}
	\begin{split}
		&\alpha J(t,\lambda^{1},y)+(1-\alpha)J(t,\lambda^2,y)\\
		=&\mathbb{E}\left[\int_t^T e^{-r(s-t)} \Big( \alpha\tilde{u}(e^{-(\rho-r)(s-t)}X_s^{*,1}) +(1-\alpha) \tilde{u}(e^{-(\rho-r)(s-t)}X_s^{*,2})\right.\\+&\left.Y_s - e^{-(\rho-r)(s-t)}(\alpha X_s^{*,1}+(1-\alpha)X_s^{*,2}) u(Y_s)\Big)ds  \right]  
		+\lambda^3 \mathbb{E}_t \left[\int_t^T e^{-\rho(s-t)}u(Y_s)ds \right]
	\end{split}
\end{eqnarray*}
By (b) in Proposition \ref{pro:dual}, the optimal processes $X^{*,j}$($j=1,2,3$) are given by 
\begin{equation*}
X_s^{*,j}=\max\left(\lambda^{j}, \sup_{t\le \xi \le s}e^{(\rho-r)(s-t)}Y_{\xi}^\gamma z^{\star}(\xi)\right).
\end{equation*}
Since $\lambda_1\neq\lambda_2$ and $X_t^{*,j}=\lambda^{j}$, we can deduce that 
$$
X_{s}^{*,1}\neq X_{s}^{*,2}\;\mbox{a.s.}
$$
Thus, by strict convexity of $\tilde{u}(\cdot)$,
$$
\alpha\tilde{u}(e^{-(\rho-r)(s-t)}X_s^{*,1}) +(1-\alpha) \tilde{u}(e^{-(\rho-r)(s-t)}X_s^{*,2})>\tilde{u}(e^{-(\rho-r)(s-t)}(\alpha X_s^{*,1}+(1-\alpha)X_s^{*,2})).
$$
Let us temporarily denote $\bar{X}^{*}=(\alpha X_s^{*,1}+(1-\alpha)X_s^{*,2})$. Then,
\begin{eqnarray*}
	\begin{split}
		&\alpha J(t,\lambda^{1},y)+(1-\alpha)J(t,\lambda^2,y)\\
		>&\mathbb{E}\left[\int_t^T e^{-r(s-t)} \Big( \tilde{u}(e^{-(\rho-r)(s-t)}\bar{X}_s^{*}) +Y_s - e^{-(\rho-r)(s-t)}\bar{X}_s^{*}u(Y_s)\Big)ds  \right]  \\
		+&\lambda^3 \mathbb{E}_t \left[\int_t^T e^{-\rho(s-t)}u(Y_s)ds \right]\\
		\ge& \inf_{X \in \mathcal{ND}(\lambda^3)} \left\{\mathbb{E}\left[\int_t^T e^{-r(s-t)} \Big( \tilde{u}(e^{-(\rho-r)(s-t)}X_s) +Y_s - e^{-(\rho-r)(s-t)}X_s u(Y_s)\Big)ds  \right] \right. \\
		+&\left. \lambda^3 \mathbb{E}_t \left[\int_t^T e^{-\rho(s-t)}u(Y_s)ds \right] \right\}\\
		=&\mathbb{E}\left[\int_t^T e^{-r(s-t)} \Big( \tilde{u}(e^{-(\rho-r)(s-t)}X_s^{*,3}) +Y_s - e^{-(\rho-r)(s-t)}X_s^{*,3} u(Y_s)\Big)ds  \right]  \\
		+&\lambda^3 \mathbb{E}_t \left[\int_t^T e^{-\rho(s-t)}u(Y_s)ds \right]=J(t,\lambda^3,y).
	\end{split}
\end{eqnarray*}
This implies that $J(t,\lambda,y)$ is strictly convex in $\lambda$ in ${\bf NR}$.
\hfill$\Box$\medskip

Now, we rewrite the Lagrangian {\bf L} in \eqref{eq:LGR5} as 
\begin{eqnarray}\label{eq:LGR6}
\begin{aligned}
{\bf L}(t,\lambda,y,X) =& \mathbb{E}_t \left[\int_t^T e^{-r(s-t))} \Big( \tilde{u}(e^{-(\rho-r)(s-t)}X_s) +Y_s - e^{-(\rho-r)(s-t)}X_s u(Y_s)\Big)ds\right] \\
+&\lambda \mathbb{E}_t \left[\int_t^T e^{-\rho(s-t)}u(Y_s)ds \right]-\lambda w. \\
\end{aligned}
\end{eqnarray}
with $X_t=\lambda, Y_t=y$. For simplicity, let ${\bf L}(t,\lambda,y,X)={\bf L}(\lambda,X)$. \\

\noindent {\bf (Step 2)} For every enforceable plan $C\in\Gamma(t,y,w)$, every $x>0$, and every $X\in\mathcal{ND}(\lambda)$, the following inequality is established:
\begin{eqnarray*}
	\begin{split}
		L(\lambda,X)\ge U_t^P(y,C)=\mathbb{E}_t\left[\int_{t}^T e^{-r(s-t)}(Y_s-C_s)ds\right]
	\end{split}
\end{eqnarray*}
The equality holds if and only if for all $s\in[t,T]$, 
$$
X_s e^{-(\rho-r)(s-t)}u'(C_s)-1=0,\;\;\int_{s}^{T}e^{-\rho(\xi-s)}(U_\xi^a(C)-U_d(\xi,Y_\xi))dX_{\xi}=0.
$$
This leads to the following {\it weakly duality} relationship:
\begin{eqnarray}
V(t,w,y)\le \inf_{\lambda>0} \left( J(t,\lambda,y)-\lambda w \right).
\end{eqnarray}
\noindent{\bf Proof of {\bf (Step 2)}} 
By the definition of $\tilde{u}(\cdot)$, 
$$
\tilde{u}(e^{-(\rho-r)(s-t)}X_s) \ge X_s e^{-(\rho-r)(s-t)}u(C_s)-C_s.
$$
This leads to 
\begin{eqnarray*}
	\begin{split}
		&{\bf L}(\lambda,X)\\\ge&\mathbb{E}_t \left[\int_t^T e^{-r(s-t)} \left(Y_s-C_s\right)ds\right] +\mathbb{E}_t\left[\int_t^T e^{-r(s-t)}\left(X_s e^{-(\rho-r)(s-t)}(u(C_s)-u(Y_s)\right)ds\right]\\
		+&\lambda \left(\mathbb{E}_t \left[\int_t^T e^{-\rho(s-t)}u(Y_s)ds \right]-w\right)\\
		=&\mathbb{E}_t\left[\int_{t}^{T} e^{-r(s-t)}(Y_s-C_s)ds\right]+\mathbb{E}_t\left[\int_{t}^{T}e^{-\rho(s-t)}\int_{s}^{T}e^{-\rho(\xi-s)}\left(u(C_{\xi})-u(Y_{\xi})\right)d\xi \cdot dX_s\right]\\
		+&\lambda \left(\mathbb{E}_t \left[\int_t^T e^{-\rho(s-t)}u(C_s)ds \right]-w\right)\\
		\ge&\mathbb{E}_t\left[\int_t^T e^{-r(s-t)}(Y_s-C_s)ds\right],
	\end{split}
\end{eqnarray*}
where the middle equation follows from integration by parts, and the last inequality follows from the fact $C$ is enforceable and $X$ is in $\mathcal{ND}(\lambda)$.

Clearly, the equality holds if and only if 
$$
X_s e^{-(\rho-r)(s-t)}u'(C_s)-1=0,\;\;\int_{s}^{T}e^{-\rho(\xi-s)}(U_\xi^a(C)-U_d(\xi,Y_\xi))dX_{\xi}=0.
$$
Moreover, we can immediately obtain 
\begin{eqnarray*}
	V(t,w,y)\le \inf_{\lambda>0} \left( J(t,\lambda,y)-\lambda w \right).
\end{eqnarray*}
\hfill$\Box$\medskip

\noindent {\bf (Step 3)} The consumption plan $C^{*}$ defined in \eqref{eq:optimal_C} is enforceable and optimal.\\

\noindent{\bf Proof of {\bf (Step 3)}} 
Let us assume that $(X_s^*)_{s=t}^{T}\in\mathcal{ND}(\lambda^*)$ minimize the Lagrangian {\bf L} in \eqref{eq:minimize_L}. Under this assumption we first prove that $(c_s^*)_{s=t}^{T}$ is enforceable and optimal. Then we will show the existence and uniqueness of $\lambda^*$ and derive $(X_s^*)_{s=t}^{T}$. Finally, we derive the optimal continuation process.

Since $(X_s^*)_{s=t}^{T}\in\mathcal{ND}(\lambda^*)$, we know that 
$$
\mathbb{E}_t\left[\int_{t}^{T}e^{-r(s-t)}|\tilde{u}(X_s^* e^{-(\rho-r)(s-t)})|\right]<\infty.
$$
From this, it is easy to check that 
$$
\mathbb{E}_t\left[\int_{t}^T e^{-r(s-t)}C_s^*ds\right]<+\infty.
$$
For sufficiently small $\delta>0$ and $h\in(0,\delta)$, we consider 
$$
\lambda^h \equiv \lambda^* + h,~~~~X_s^h \equiv X_s^* + h,~~~ \textrm{and}~~~X_s^\delta \equiv X_s^* + \delta
$$
Then
\begin{eqnarray}\label{eq:enforceable1}
\begin{split}
&\mathbb{E}_t\left[\int_{t}^{T}e^{-r(s-t)}|\tilde{u}((X_s^*+\delta) e^{-(\rho-r)(s-t)})|ds\right]\\
&<\left|\dfrac{\gamma}{1-\gamma}\right|\mathbb{E}_t\left[\int_{t}^{T}e^{-r(s-t)}\left(e^{-(\rho-r)(s-t)}(X_s^*(1+\delta/X_t^*))\right)^{\frac{1}{\gamma}}ds\right]\\
&<(1+\delta/\lambda)^{\frac{1}{\gamma}}\mathbb{E}_t\left[\int_{t}^{T}e^{-r(s-t)}|\tilde{u}(X_s^* e^{-(\rho-r)(s-t)})|\right]<\infty.
\end{split}
\end{eqnarray}
Similarly,
\begin{eqnarray}\label{eq:enforceable15}
\begin{split}
\mathbb{E}_t\left[\int_{t}^{T}e^{-r(s-t)}|\tilde{u}((X_s^*(1\pm\delta))e^{-(\rho-r)(s-t)})|ds\right]<\infty.
\end{split}
\end{eqnarray}
Since 
$$
\mathbb{E}_t\left[\int_{t}^{T}e^{-\rho(s-t)}|U_d(Y_s)|X_s^* ds\right]<\infty,
$$
clearly, we can have 
$$
\mathbb{E}_t\left[\int_{t}^{T}e^{-\rho(s-t)}|U_d(Y_s)|(X_s+\delta) ds\right]<\infty.
$$
{The convexity of $\tilde{u}(\cdot)$} implies 
\begin{eqnarray}
\begin{split}\label{eq:enforceable2}
e^{-(\rho-r)(s-t)}u(C_s^*)\le& \dfrac{\tilde{u}(X_s^h e^{-(\rho-r)(s-t)})-\tilde{u}(X_s^* e^{-(\rho-r)(s-t)})}{h}\\\le& \dfrac{\tilde{u}(X_s^\delta e^{-(\rho-r)(s-t)})-\tilde{u}(X_s^* e^{-(\rho-r)(s-t)})}{\delta}
\end{split}
\end{eqnarray}
Thus, 
\begin{eqnarray}\label{eq:enforceable3}
\begin{split}
&\mathbb{E}\left[\int_t^{T}e^{-\rho(s-t)}|u(C_s^*)|ds\right]\\
\le&\mathbb{E}\left[\int_t^{T}e^{-r(s-t)}|u(C_s^*)|\dfrac{e^{-(\rho-r)(s-t)}X_s^*}{\lambda^*}ds\right]\\
\le &\dfrac{1}{\lambda^*}\mathbb{E}\left[\int_t^{T}e^{-r(s-t)}\left(|\tilde{u}(X_s^*e^{-(\rho-r)(s-t)})|+C_s^*\right)ds\right]<\infty.
\end{split}
\end{eqnarray}
\eqref{eq:enforceable1},\eqref{eq:enforceable2}, and \eqref{eq:enforceable3} imply $X^h \in \mathcal{ND}(\lambda^* + h)$.

Since $(X_s^*)_{s=t}^{T}\in\mathcal{ND}(\lambda^*)$ minimizes the Lagrangian {\bf L},
$$
{\bf L}(\lambda^h, X^h) \ge {\bf L}(\lambda^*, X^*).
$$
Hence, 
$$
\lim_{h \downarrow 0} \dfrac{{\bf L}(\lambda^h,X^h)-{\bf L}(\lambda^*,X^*)}{h}\ge 0.
$$
or, equivalently, 
$$
\lim_{h \downarrow 0}\mathbb{E}_t\left[\int_t^T e^{-r(s-t)}\dfrac{\tilde{u}(X_s^h e^{-(\rho-r)(s-t)})-\tilde{u}(X_s^* e^{-(\rho-r)(s-t)})}{h}ds\right] -w \ge 0.
$$
The Dominated Convergence Theorem implies 
\begin{eqnarray*}
	\begin{split}
		\lim_{h \downarrow 0}\;&\mathbb{E}_t\left[\int_t^T e^{-r(s-t)}\dfrac{\tilde{u}(X_s^h e^{-(\rho-r)(s-t)})-\tilde{u}(X_s^* e^{-(\rho-r)(s-t)})}{h}ds\right] -w\\
		=&\mathbb{E}_t\left[\lim_{h \downarrow 0}\int_t^T e^{-r(s-t)}\dfrac{\tilde{u}(X_s^h e^{-(\rho-r)(s-t)})-\tilde{u}(X_s^* e^{-(\rho-r)(s-t)})}{h}ds\right] -w \\
		=&\mathbb{E}_t\left[\int_t^T e^{-\rho(s-t)}u(C_s^*)ds\right]-w \ge 0.
	\end{split}
\end{eqnarray*}
Thus, $C^*$ satisfies the promise-keeping constraints.

Similar to \citet{MJ}, define $X^h(w,\xi)\equiv X^*(t,\xi)+h{\bf 1}_{A\times(s,T]}(w,\xi)$ for $h\in(0,\delta)$, $s\in[t,T]$ and $A\in\mathcal{F}_s$. Note $X_t^h=X_t^*=\lambda^*$. 

By a similar argument, we can obtain 
$$
X^{h}\in\mathcal{ND}(\lambda^*).
$$
Since 
$$
\lim_{h \downarrow 0}\dfrac{{\bf L}(\lambda^*,X^h)-{\bf L}(\lambda^*,X^*)}{h}\ge 0.
$$
This leads to 
$$
\mathbb{E}_t\left[{\bf 1}_{A}\int_{s}^{T}e^{-\rho(\xi-t)}U(C_\xi^*)d\xi\right]\ge \mathbb{E}_t\left[{\bf 1}_{A}e^{-\rho(s-t)}U_d(s,Y_s)\right].
$$
Since $A$ was an arbitrary set in $\mathcal{F}_s$, we deduce that 
\begin{eqnarray*}
	\begin{split}
		U_s(C)=\mathbb{E}_s\left[\int_{s}^{T}e^{-\rho(\xi-s)}U(C_\xi^*)d\xi\right]\ge U_d(s,Y_s),
	\end{split}
\end{eqnarray*}
for any $s\in[t,T]$.

Therefore, the consumption plan $C^*$ is enforceable. Now we will show that $C^*$ is optimal.

For $h\in(0,\delta)$, let us consider 
$$
X^{\pm h}\equiv X^*(1\pm h).
$$
By the following convexity of $\tilde{u}(\cdot)$, 
\begin{eqnarray*}
	\begin{split}
		\dfrac{\tilde{u}(X_s^{-\delta}e^{-(\rho-r)(s-t)})-\tilde{u}(X_s^{*}e^{-(\rho-r)(s-t)})}{-\delta}
		\le&\dfrac{\tilde{u}(X_s^{\pm h}e^{-(\rho-r)(s-t)})-\tilde{u}(X_s^{*}e^{-(\rho-r)(s-t)})}{\pm h}\\
		\le&\dfrac{\tilde{u}(X_s^{\delta}e^{-(\rho-r)(s-t)})-\tilde{u}(X_s^{*}e^{-(\rho-r)(s-t)})}{\delta}
	\end{split}
\end{eqnarray*}
and the condition \eqref{eq:enforceable15}, we can deduce $X^{\pm h}\in\mathcal{ND}(\lambda^*(1+h))$. 

Since ${\bf L}(\lambda^*(1\pm h),X^{\pm h})\ge {\bf L}(\lambda^*,X^{*})$,
$$
\lim_{h\downarrow 0}\dfrac{{\bf L}(\lambda^*(1+h),X^{h})-{\bf L}(\lambda^*,X^{*})}{h}\ge 0,\;\;\lim_{h\uparrow 0}\dfrac{{\bf L}(\lambda^*(1-h),X^{-h})-{\bf L}(\lambda^*,X^{*})}{-h}\ge 0
$$
By the Dominated Convergence theorem and integration by parts,
$$
\lambda^*\left(U_t(C^*)-w\right)+\mathbb{E}_t\left[\int_{t}^{T}e^{-\rho(s-t)}\int_{s}^{T}e^{-\rho(\xi-s)}\left(u(C_{\xi}^*)-u(Y_{\xi})\right)d\xi \cdot dX_s\right]=0.
$$
Thus, the promise keeping constraint and the participation constraint must hold with equality for the consumption plan $C^*$.

Since 
$$
U_t^P(y,C^*)\le\sup_{C \in \Gamma(t,y,w)}U_t^P(t,y,C)\le \inf_{X \in \mathcal{ND}(\lambda)}{\bf L}(t,\lambda,y,X)-\lambda w \le{\bf L}(t,\lambda^*,X^*)-\lambda^* w,
$$
we conclude that the consumption plan $C^*$ is optimal and the following duality relationship holds:
$$
V(t,w,y)=\inf_{\lambda>0}\left\{J(t,\lambda,y)-\lambda w\right\}.
$$
\hfill$\Box$\medskip

\noindent {\bf (Step 4)} Determination of $\lambda^*$ in the duality relationship \eqref{eq:duality} and the optimal process $(X_s^*)_{s=t}^{T}$.\\

By Lemma \ref{lem:integral_Q} and \eqref{eq:partial_J}, 
\begin{eqnarray*}
	\begin{split}
		\partial_{\lambda}J(t,\lambda,y)=&y^{1-\gamma}g(t,\dfrac{\lambda}{y^{\gamma}})+\partial_{\lambda}J_0(t,\lambda,y)\\
		=&\dfrac{y^{1-\gamma}}{1-\gamma}\int_{t}^{T}e^{-\hat{\rho}(s-t)}\mathcal{N}\left(-d^1(s-t,\dfrac{\lambda}{y^{\gamma}z^{\star}(s)})\right)ds\\~+&\dfrac{\lambda^{\frac{1}{\gamma}-1}}{1-\gamma}\int_{t}^{T}e^{-K(s-t)}\mathcal{N}\left(d^{\gamma}(s-t,\dfrac{\lambda}{y^\gamma z^\star(s)})\right)ds.\\
	\end{split}
\end{eqnarray*}
and we deduce that 
\begin{eqnarray*}
	\begin{split}
		\lim_{\lambda \to \infty}\partial_{\lambda}J(t,\lambda,y)=\begin{cases}
			\infty\qquad&\mbox{if}\;\;0<\gamma<1,\\
			0\qquad&\mbox{if}\;\;1<\gamma.
		\end{cases}
	\end{split}
\end{eqnarray*}
Moreover, by Lemma \ref{lem-cinfinity} or the value matching condition for $g(t,z)$,
$$
\partial_{\lambda}J(t,z^{\star}(t)y^\gamma,y)=U_d(t,y).
$$
By {\bf (Step 1)}, $J(t,\lambda,y)$ is strictly convex in $\lambda$ in {\bf NR}, i.e.,
$$
\partial_{\lambda\lambda}J(t,\lambda,y)>0 \;\;\mbox{in}\;\;{\bf NR}.
$$
This implies that $\partial_{\lambda}J(t,\lambda,y)$ is strictly increasing in $\lambda$ in {\bf NR}. 

Thus, for given $w$ satisfying Assumption \ref{as:promise_value}, there exists a unique $\lambda^*$ such that 
$$
w=\partial_{\lambda}J(t, {\lambda^*},y)
$$
and $(t,\lambda^*,y)\in{\bf NR}$.

Now, we will determine the optimal shadow process $(X_s^*)_{s=t}^T\in\mathcal{ND}(\lambda^*)$.

{Since the optimal stopping time $\tau$ in Problem \ref{pr:OS} is given by
	\begin{eqnarray*}
		\tau = \inf\{s\mid 1\le {e^{(\rho-r)(s-t)}}z^\star(s){Y_s^\gamma}\},
	\end{eqnarray*}
	the optimal stopping time $\tau(x)$ in Lemma \ref{thm:lem1} can be written by 
	\begin{eqnarray*}
		\tau(x) = \inf\{s\mid x\le {e^{(\rho-r)(s-t)}}z^\star(s){Y_s^\gamma}\}.
	\end{eqnarray*}
}
Since 
$$
\{X_s^*<x\}=\{s<\tau(x)\}=\left\{\max_{t\le \t \le s}e^{(\rho-r)(\t-t)}z^{\star}(\t)Y_\t^\gamma <x\right\},
$$
the optimal process $(X_s^*)_{s=t}^T$ can be expressed explicitly as 
\begin{eqnarray*}
	X_s^*=\max\left(\lambda^*, \max_{t\le \t \le s}e^{(\rho-r)(\t-t)}z^{\star}(\t)Y_\t^\gamma \right).
\end{eqnarray*}

By defining $\lambda_s^*=e^{-(\rho-r)(s-t)}X_s^*$, since Problem \ref{pro:dual} is {\it time consistent}, we can deduce that dual value function $J(s,\lambda_s^*,Y_s)$ at time $s\in[t,T]$ is given by 
\begin{eqnarray*}
	\begin{split}
		J(s,\lambda_s^*,Y_s)=&-\dfrac{Y_s^{1-\gamma}}{1-\gamma}\int_{\lambda_s^*}^{\infty}\left[\int_{s}^{T}e^{-\hat{\rho}(\xi-s)}\mathcal{N}\left(-d^1(\xi-s,\frac{x}{ z^{\star}(\xi)Y_s^\gamma})\right)d\xi\right.\\&\left.-\left(\dfrac{x}{Y_s^\gamma}\right)^{\frac{1}{\gamma}-1}\int_{s}^{T}e^{-\hat{\rho}(\xi-s)}\mathcal{N}\left(-d^1(\xi-s,\frac{x}{z^{\star}(\xi)Y_s^\gamma})\right)d\xi\right]dx\\
		&+\dfrac{\gamma}{1-\gamma}\left(\dfrac{1-e^{-K(T-s)}}{K}(\lambda_s^*)^{\frac{1}{\gamma}}+\dfrac{1-e^{-\hat{r}(T-s)}}{\hat{r}}Y_s\right).
	\end{split}
\end{eqnarray*}
and satisfies the duality-relationship:
\begin{eqnarray*}
	{V(s,Y_s,w_s)=\inf_{\lambda>0}\left\{J(s,\lambda,Y_s) - w_s \lambda\right\}=J(s,\lambda_s^*,Y_s) - w_s \lambda_s^*.}
\end{eqnarray*}
By the first-order condition, we obtain
\begin{eqnarray*}
	\begin{split}
		w_s =&\partial_{\lambda}J(s,\lambda_s^*,Y_s)\\
		=&\dfrac{Y_s^{1-\gamma}}{1-\gamma}\int_{t}^{T}e^{-\hat{\rho}(s-t)}\mathcal{N}\left(-d^1(s-t,\dfrac{\lambda_s^*}{Y_s^{\gamma}z^{\star}(s)})\right)ds\\+&\dfrac{(\lambda_s^*)^{\frac{1}{\gamma}-1}}{1-\gamma}\int_{t}^{T}e^{-K(s-t)}\mathcal{N}\left(d^{\gamma}(s-t,\dfrac{\lambda_s^*}{Y_s^\gamma z^\star(s)})\right)ds.
	\end{split}
\end{eqnarray*}
\noindent {\bf (Step 5)} The optimal process $(X_s^*)_{s=t}^T$ satisfies the integrability condition \eqref{eq:integrability}.\\

\noindent{\bf Proof of {\bf (Step 5)}} 
It is enough to show that 
$$
\mathbb{E}_t\left[\int_t^T e^{-r(s-t)}(e^{-(\rho-r)(s-t)}X_s^*)^{\frac{1}{\gamma}} ds\right]<\infty\;\;\;\mbox{and}\;\;\;\mathbb{E}_t\left[\int_t^T e^{-\rho(s-t)}Y_s^{1-\gamma}X_s^*ds\right]<\infty
$$
First, we will utilize the idea in Lemma \ref{thm:lem1}.
\begin{eqnarray*}
	\begin{split}
		&\mathbb{E}_t\left[\int_t^T e^{-r(s-t)}(e^{-(\rho-r)(s-t)}X_s^*)^{\frac{1}{\gamma}} ds\right]\\
		=&\mathbb{E}_t\left[\int_t^T e^{-K(s-t)}(X_s^*)^\frac{1}{\gamma}ds\right]\\
		=&\mathbb{E}_t\left[\int_t^T e^{-K(s-t)}\left(\int_{X_t^*}^{X_s^*}\frac{1}{\gamma}x^{\frac{1}{\gamma}-1}dx + X_t^*\right)ds\right]\\
		=&\dfrac{1-e^{-K(T-t)}}{K}\lambda^* + \dfrac{1}{\gamma}\mathbb{E}_t\left[\int_t^T e^{-K(s-t)}\left(\int_{\lambda^*}^{\infty}x^{\frac{1}{\gamma}-1}{\bf 1}_{\{X_s^*\ge x\}}dx\right)ds\right]\\
		=&\dfrac{1-e^{-K(T-t)}}{K}\lambda^* +\dfrac{1}{\gamma}\int_{\lambda^*}^{\infty}\int_t^T e^{-K(s-t)}x^{\frac{1}{\gamma}-1}\mathbb{E}_t\left[{\bf 1}_{\{s\ge \tau(x)^*\}}\right]ds dx,
	\end{split}
\end{eqnarray*}
where the last equality is obtained from Fubini's theorem.

By Proposition \ref{pro:dual}, we can obtain 
$$
\mathbb{P}(s\ge \tau(x)^*)=\mathbb{P}(x\mathcal{H}_s \le z^\star(s)).
$$
Since 
$$
d\mathcal{H}_s = (\hat{r}-\hat{\rho}+\gamma\sigma^2)\mathcal{H}_s ds -\gamma\sigma\mathcal{H}_s dB_s
$$
under the measure $\mathbb{P}$ with $\mathcal{H}_t= 1/y^\gamma$, we can easily obtain that 
$$
\mathbb{P}(s\ge \tau(x)^*)=\mathcal{N}\left(-d^\gamma(s-t,\dfrac{x}{z^\star(s)y^\gamma})\right).
$$
For $\widetilde{z}^\star(s)=z^\star(T-s)$, $\xi=s-t$ and $\tau=T-t$,
\begin{eqnarray}\label{eq:integral1}
\begin{split}
&\mathbb{E}_t\left[\int_t^T e^{-r(s-t)}(e^{-(\rho-r)(s-t)}X_s^*)^{\frac{1}{\gamma}} ds\right]\\
=&\dfrac{1-e^{-K(T-t)}}{K}\lambda^* + \dfrac{1}{\gamma}\int_{\lambda^*}^{\infty}\int_0^\tau e^{-K\xi}x^{\frac{1}{\gamma}-1}\mathcal{N}\left(-d^\gamma(\xi,\dfrac{x}{\widetilde{z}^\star(\tau-\xi)y^\gamma})\right)d\xi dx\\
< &\dfrac{\lambda^*}{K} + \dfrac{1}{\gamma}\int_{\lambda^*}^{\infty}\lim_{\tau \to \infty}\left[\int_0^\tau e^{-K\xi}x^{\frac{1}{\gamma}-1}\mathcal{N}\left(-d^\gamma(\xi,\dfrac{x}{\widetilde{z}^\star(\tau-\xi)y^\gamma})\right)d\xi \right]dx\\
=&\dfrac{\lambda^*}{K} + \dfrac{1}{\gamma}\int_{\lambda^*}^{\infty}x^{\frac{1}{\gamma}-1}\int_0^\infty e^{-K\xi}\mathcal{N}\left(-d^\gamma(\xi,\dfrac{x}{z_\infty y^\gamma})\right)d\xi dx.
\end{split}
\end{eqnarray}
By using the result in Proof of Lemma \ref{lem:inf}, we can derive  
\begin{eqnarray}
\begin{split}\label{eq:integral2}
\int_0^\infty e^{-K\xi}\mathcal{N}\left(-d^\gamma(\xi,\dfrac{x}{z_\infty y^\gamma})\right)d\xi=\frac{1}{K}\left(\frac{\alpha_++\left(\frac{\gamma-1}{\gamma}\right)}{\alpha_+-\alpha_-}\right)\cdot \left(\frac{x}{z_\infty y^\gamma}\right)^{\alpha_-+\left(\frac{\gamma-1}{\gamma}\right)}
\end{split}
\end{eqnarray}
By \eqref{eq:integral1} and \eqref{eq:integral2}, we deduce 
\begin{eqnarray*}
	\begin{split}
		&\mathbb{E}_t\left[\int_t^T e^{-r(s-t)}(e^{-(\rho-r)(s-t)}X_s^*)^{\frac{1}{\gamma}} ds\right]\\
		<&\dfrac{\lambda^*}{K}+\dfrac{1}{\gamma}\dfrac{1}{K}\left(\frac{\alpha_++\left(\frac{\gamma-1}{\gamma}\right)}{\alpha_+-\alpha_-}\right)(z_\infty y^\gamma)^{\frac{1}{\gamma}-1-\alpha_-}\int_{\lambda^*}^{\infty}x^{\alpha_-}dx<\infty.
	\end{split}
\end{eqnarray*}
Similarly,
\begin{eqnarray*}
	\begin{split}
		&\mathbb{E}_t\left[\int_t^T e^{-\rho(s-t)}Y_s^{1-\gamma}X_s^*ds\right]\\
		=&\mathbb{E}_t\left[\int_t^T e^{-\rho(s-t)}Y_s^{1-\gamma}\left(\int_{X_t^*}^{X_s^*}1dx + X_t^*\right)\right]\\
		=&\dfrac{1-e^{-\hat{\rho}(T-t)}}{\hat{\rho}}y^{1-\gamma}\lambda^* + \mathbb{E}_t \left[\int_t^T e^{-{\rho}(s-t)}Y_s^{1-\gamma}\left(\int_{X_t^*}^{X_s^*}1dx\right) ds\right]\\
		=&\dfrac{1-e^{-\hat{\rho}(T-t)}}{\hat{\rho}}y^{1-\gamma}\lambda^*+y^{1-\gamma}\mathbb{E}^{\mathbb{Q}}\left[\int_t^T e^{-\hat{\rho}(s-t)}\left(\int_{\lambda^*}^{\infty}{\bf 1}_{\{X_s^* \ge x \}}\right)ds\right]\\
		=&\dfrac{1-e^{-\hat{\rho}(T-t)}}{\hat{\rho}}y^{1-\gamma}\lambda^* + y^{1-\gamma}\int_{\lambda^*}^{\infty}\int_t^Te^{-\hat{\rho}(s-t)}\mathbb{E}^{\mathbb{Q}}\left[{\bf 1}_{\{s\ge \tau(x)^* \}}\right]ds dx,
	\end{split}
\end{eqnarray*}
where the measure $\mathbb{Q}$ is defined in Lemma \ref{thm:lem1}.

Since 
$$
d\mathcal{H}_s = (\hat{r}-\hat{\rho}+\gamma^2\sigma^2)\mathcal{H}_s ds -\gamma\sigma\mathcal{H}_s dB_s^{\mathbb{Q}},
$$
we have 
$$
\mathbb{Q}(s\ge \tau(x)^*)=\mathbb{Q}(x\mathcal{H}_s \le z^{\star}(t))=\mathcal{N}\left(-d^1(s-t,\dfrac{x}{z^\star(s)y^\gamma})\right).
$$
By using the result in Proof of Lemma \ref{lem:inf}, we deduce that 
\begin{eqnarray*}
	\begin{split}
		&\mathbb{E}_t\left[\int_t^T e^{-\rho(s-t)}Y_s^{1-\gamma}X_s^*ds\right]\\
		=&\dfrac{1-e^{-\hat{\rho}(T-t)}}{\hat{\rho}}y^{1-\gamma}\lambda^* + y^{1-\gamma}\int_{\lambda^*}^{\infty}\int_t^Te^{-\hat{\rho}(s-t)}\mathcal{N}\left(-d^1(s-t,\dfrac{x}{z^\star(s)y^\gamma})\right)ds dx\\
		<&\dfrac{y^{1-\gamma}\lambda^*}{\hat{\rho}}+y^{1-\gamma}\int_{\lambda^*}^{\infty}\int_0^{\infty}e^{-\hat{\rho}\xi}\mathcal{N}\left(-d^1(\xi,\dfrac{x}{z_\infty y^\gamma})\right)d\xi dx\\
		=&\dfrac{y^{1-\gamma}\lambda^*}{\hat{\rho}}+\dfrac{1}{\hat{\rho}}\left(\dfrac{\alpha_+}{\alpha_+ - \alpha_-}\right)(z_\infty)^{-\alpha_-}y^{1-\gamma-\alpha_-\gamma}\int_{\lambda^*}^\infty x^{\alpha_-}dx <\infty.
	\end{split}
\end{eqnarray*}
Thus, we can conclude that the optimal process $(X_s^*)_{s=t}^T$ satisfies the integrability condition \eqref{eq:integrability}. \hfill$\Box$\medskip

From {\bf (Step 1)}--{\bf (Step 5)}, we have just proved Theorem \ref{thm:main}.

\hfill$\Box$\medskip

We will now show the convergence of the optimal strategies for a finite horizon problem to those for the infinite horizon problem.
\begin{thm}\label{thm:infinite}~
	For $t\ge 0$, as time to maturity goes to infinity, i.e., $T-t\to \infty$, the agent's optimal consumption $C_{t,\infty}^*$ and the optimal promised value $w_{t,\infty}^*$ at time $t$ are given by 
	\begin{eqnarray*}
		\begin{split}
			C_{t,\infty}^*=&(\lambda_{t,\infty}^*)^{\frac{1}{\gamma}},\\
			w_{t,\infty}^*=&-\dfrac{1}{\gamma}\dfrac{1}{K}\dfrac{1}{\alpha_-}(z_\infty Y_t^\gamma)^{\frac{1}{\gamma}-1-\alpha_-}(\lambda_{t,\infty}^*)^{\alpha_-} + \dfrac{1}{1-\gamma}\dfrac{1}{K}(\lambda_{t,\infty}^*)^{\frac{1}{\gamma}-1},
		\end{split}
	\end{eqnarray*}
	where $\alpha_-$ and  $z_\infty$ are defined in \eqref{eq:f1} and \eqref{eq:INFZ} in Section \ref{sec:B}, respectively,  $\lambda_{t,\infty}^*=e^{-(\rho-r)t}X_{t,\infty}^*$, and 
	$$ 
	X_{t,\infty}^*=\max\left(\lambda_{\infty}^*, z_\infty\sup_{0\le \xi \le t}e^{(\rho-r)\xi}Y_{\xi}^\gamma \right).
	$$
	Moreover, $\lambda_{\infty}^*$ is a unique solution to the following algebraic equation:
	\begin{eqnarray*}
		\begin{split}
			w=&-\dfrac{1}{\gamma}\dfrac{1}{K}\dfrac{1}{\alpha_-}(z_\infty y^\gamma)^{\frac{1}{\gamma}-1-\alpha_-}(\lambda_{\infty}^*)^{\alpha_-} + \dfrac{1}{1-\gamma}\dfrac{1}{K}(\lambda_{\infty}^*)^{\frac{1}{\gamma}-1}.
		\end{split}
	\end{eqnarray*}
\end{thm}
\noindent{\bf Proof.} 
By Lemma \ref{lem:inf}, 
$$
\lim_{T-t \to \infty} Q(t,z)=Q_\infty (z)\;\;\;\mbox{and}\;\;\;\lim_{T-t\to \infty}z^\star(t)=z_\infty,
$$
where $Q_\infty(z)$ and $z_\infty$ are defined in \eqref{eq:INFQ} and \eqref{eq:INFZ}, respectively.

Thus, we have 
\begin{eqnarray*}
	\begin{split}
		g_\infty(z)\equiv \lim_{T-t \rightarrow \infty} g(t,z)=Q_\infty(z)+\dfrac{1}{1-\gamma}\left(\dfrac{1}{\hat{\rho}}-\dfrac{1}{K}z^{\frac{1}{\gamma}-1}\right),
	\end{split}
\end{eqnarray*}
and 
\begin{eqnarray*}
	\begin{split}
		J_\infty(y,w)=\lim_{T-t \rightarrow \infty}J(t,y,w)=-y^{1-\gamma}\int_{\lambda}^{\infty}g_\infty\left(\dfrac{x}{y^\gamma}\right)dx +\dfrac{\gamma}{1-\gamma}\dfrac{1}{K}\lambda^{\frac{1}{\gamma}} +\dfrac{1}{\hat{r}}y.
	\end{split}
\end{eqnarray*}
From Theorem \ref{thm:main}, we can derive 
\begin{eqnarray*}
	\begin{split}
		C_{t,\infty}^*=&(\lambda_{t,\infty}^*)^{\frac{1}{\gamma}}\\
		w_{t,\infty}^*=&\partial_\lambda J_\infty(\lambda_{t,\infty}^*Y_t^\gamma)\\
		=&-\dfrac{1}{\gamma}\dfrac{1}{K}\dfrac{1}{\alpha_-}(z_\infty Y_t^\gamma)^{\frac{1}{\gamma}-1-\alpha_-}(\lambda_{t,\infty}^*)^{\alpha_-} + \dfrac{1}{1-\gamma}\dfrac{1}{K}(\lambda_{t,\infty}^*)^{\frac{1}{\gamma}-1}.
	\end{split}
\end{eqnarray*}
where $\lambda_{t,\infty}^*=e^{-(\rho-r)t}X_{t,\infty}^*$ and 
$$ 
X_{t,\infty}^*=\max\left(\lambda_{\infty}^*, z_\infty\sup_{0\le \xi \le t}e^{(\rho-r)\xi}Y_{\xi}^\gamma \right).
$$
Here, $\lambda_{\infty}^*$ is a unique solution to the following algebraic equation:
\begin{eqnarray*}
	\begin{split}
		w=&-\dfrac{1}{\gamma}\dfrac{1}{K}\dfrac{1}{\alpha_-}(z_\infty y^\gamma)^{\frac{1}{\gamma}-1-\alpha_-}(\lambda_{\infty}^*)^{\alpha_-} + \dfrac{1}{1-\gamma}\dfrac{1}{K}(\lambda_{\infty}^*)^{\frac{1}{\gamma}-1}.
	\end{split}
\end{eqnarray*}
For this infinite horizon case, the uniqueness of $\lambda_{\infty}^*$ was proven by \cite{MJ}.
\hfill$\Box$\medskip

\begin{rem}~
	For $\alpha=1-\gamma$ and $\beta=1-\gamma-\gamma\alpha_-$, the results in Theorem \ref{thm:infinite}  {are} consistent  {with} those of Section 2 in \citet{MJ}.
\end{rem}
\section{Numerical Illustrations}\label{sec:4}
In this section, we illustrate numerical simulation results for optimal contracting policies. The optimal contracting policies stated in Theorem \ref{thm:main} is not fully explicit, since it requires solving the integral equation \eqref{eq:integral} for the free boundary $z^{\star}(t)$. Thus, we should solve for the free boundary by numerical methods. Thus, we solve the integral equation \eqref{eq:integral} by the recursive integration method proposed by \citet{Huang}.
\begin{figure}[h]
	\centering
	\subfigure[$\lambda^*/Y^\gamma$]{\label{fig00b}\includegraphics[scale=0.45]{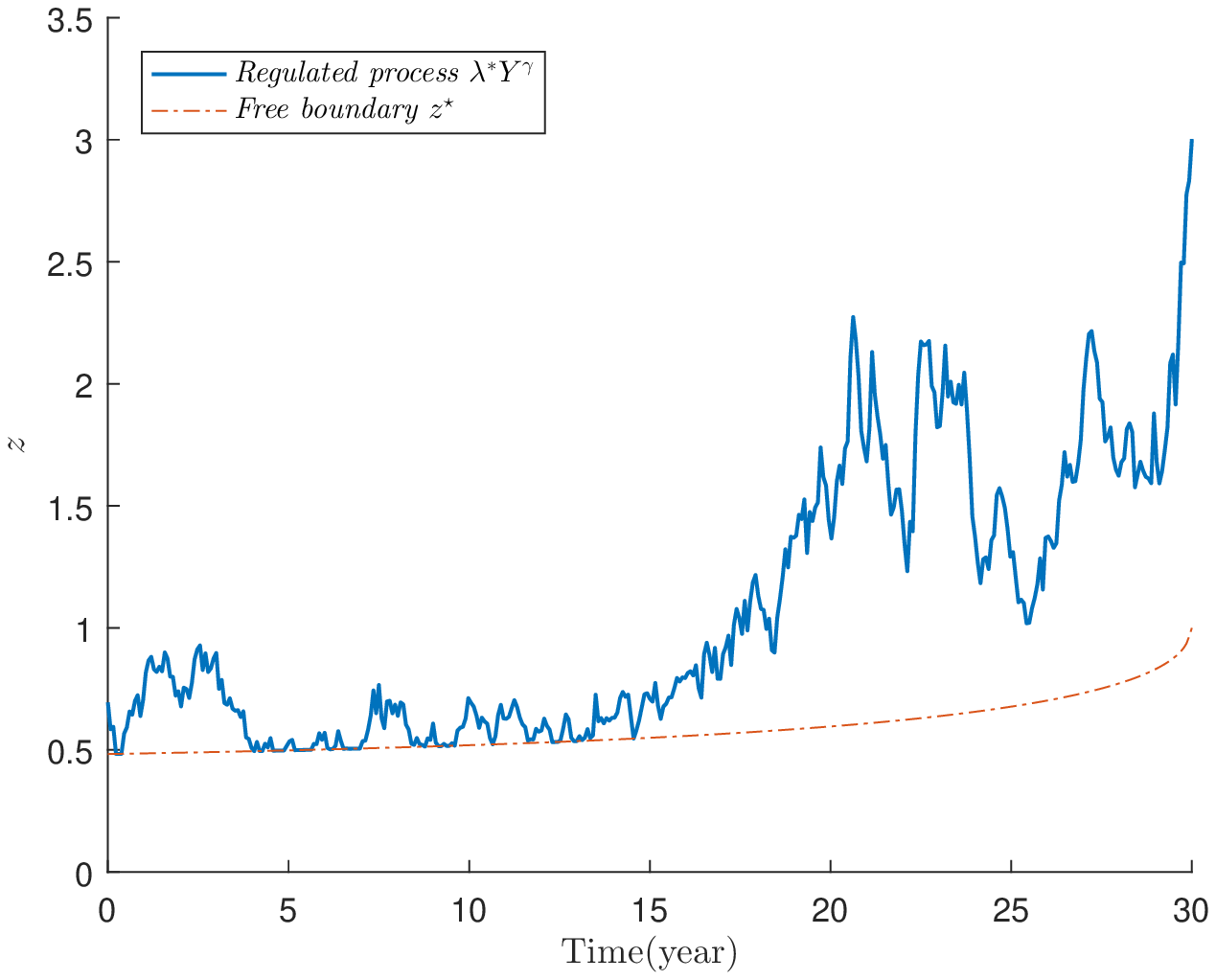}}
	\subfigure[$X^*$]{\label{fig00b}\includegraphics[scale=0.45]{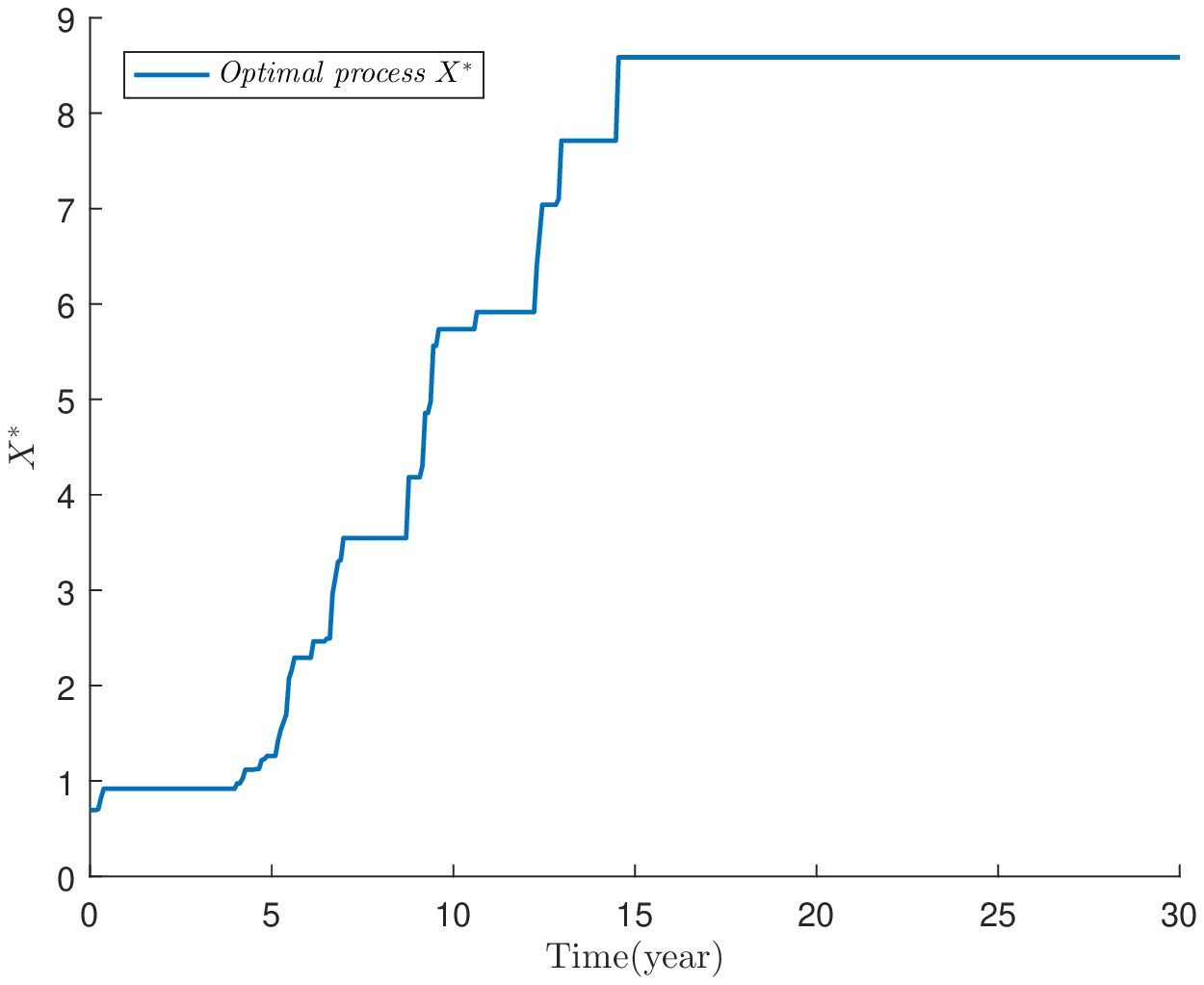}}
	\subfigure[$C^*$]{\label{fig00a}\includegraphics[scale=0.45]{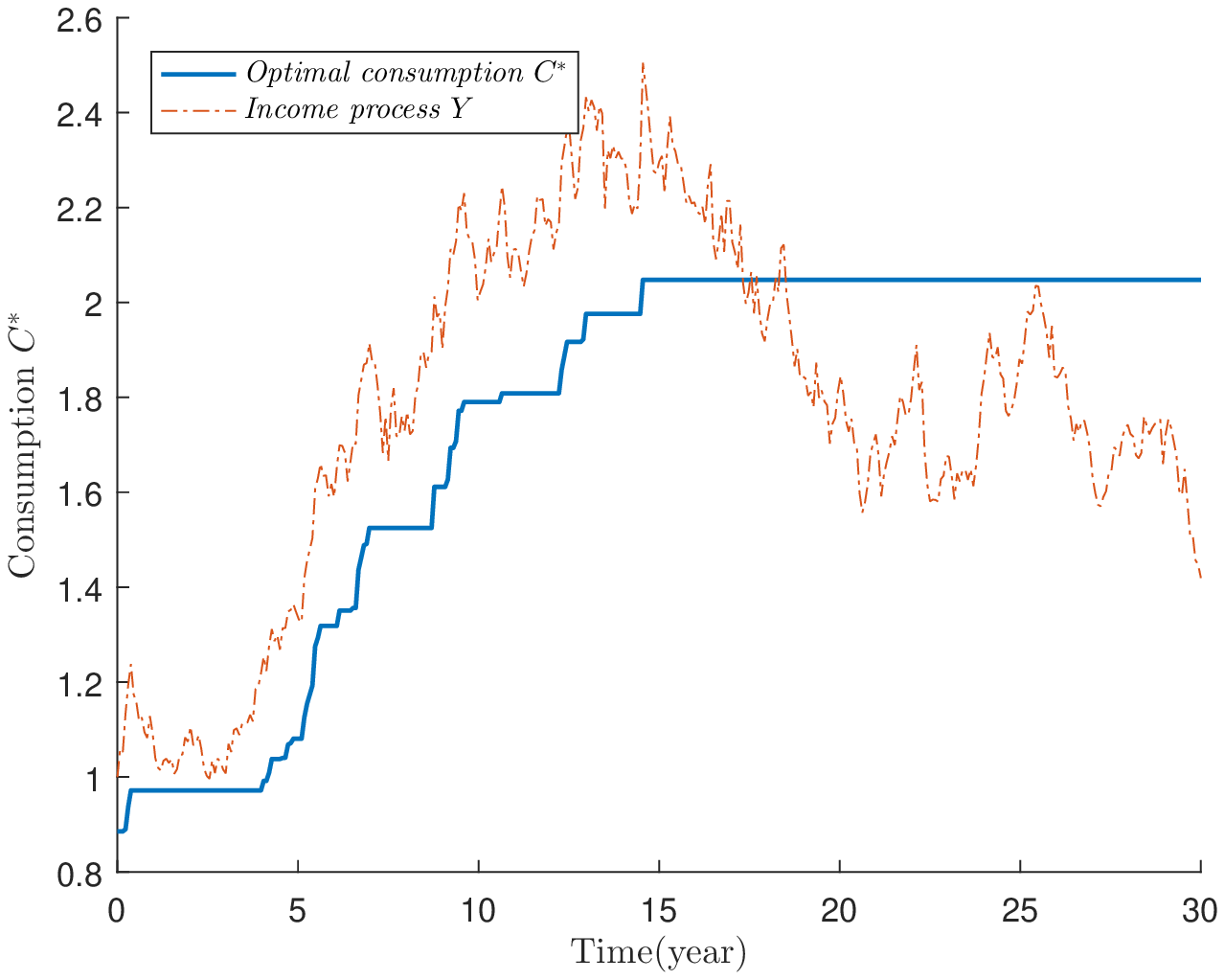}}
	\caption{\label{fig:2} Simulation of the optimal consumption $C^*$, the optimal process $X^*$ and the regulated process $\lambda^*/Y^\gamma$. The parameter values are as follows: $\rho=0.04,\;r=0.04,\;\mu=0.02\;\sigma=0.1,\;\gamma=3,\;w=-5$, and $T=30$.}
\end{figure}

Figure \ref{fig:2} presents simulation paths of the optimal consumption $C^*$, the optimal costate process $X^*$, and the regulated process ${\lambda^*}/{Y^\gamma}$. The optimal costate process $X_s^*$ ensures that $(s,{\lambda_s^*}/{Y_s^\gamma})$ will never leave the no-jump region for $s\in[t,T]$ as shown in Figure \ref{fig:2} (a). Whenever $\lambda^*/Y^\gamma$ low enough to hit the free boundary $z^\star$, the non-decreasing process $X^*$ increases in Figure \ref{fig:2} (b). The reason is that the optimal costate process $X^*$ have the property that it increases only when $\lambda^*/Y^\gamma$ hits the free boundary, at which time the participation constraints \eqref{eq:parti} bind. This leads to increase {s in} the optimal consumption $C^*$. 
In other words, if income process $Y$ increases enough and thus $\lambda^*/Y^\gamma$ decreases and hits the free boundary, the principal should increase the agent's consumption $C^*$ so that the agent does not walk away from the contract{,} as shown in Figure \ref{fig:2} (c).
\begin{figure}[h]
	\centering
	\subfigure{\label{fig01a}\includegraphics[scale=0.45]{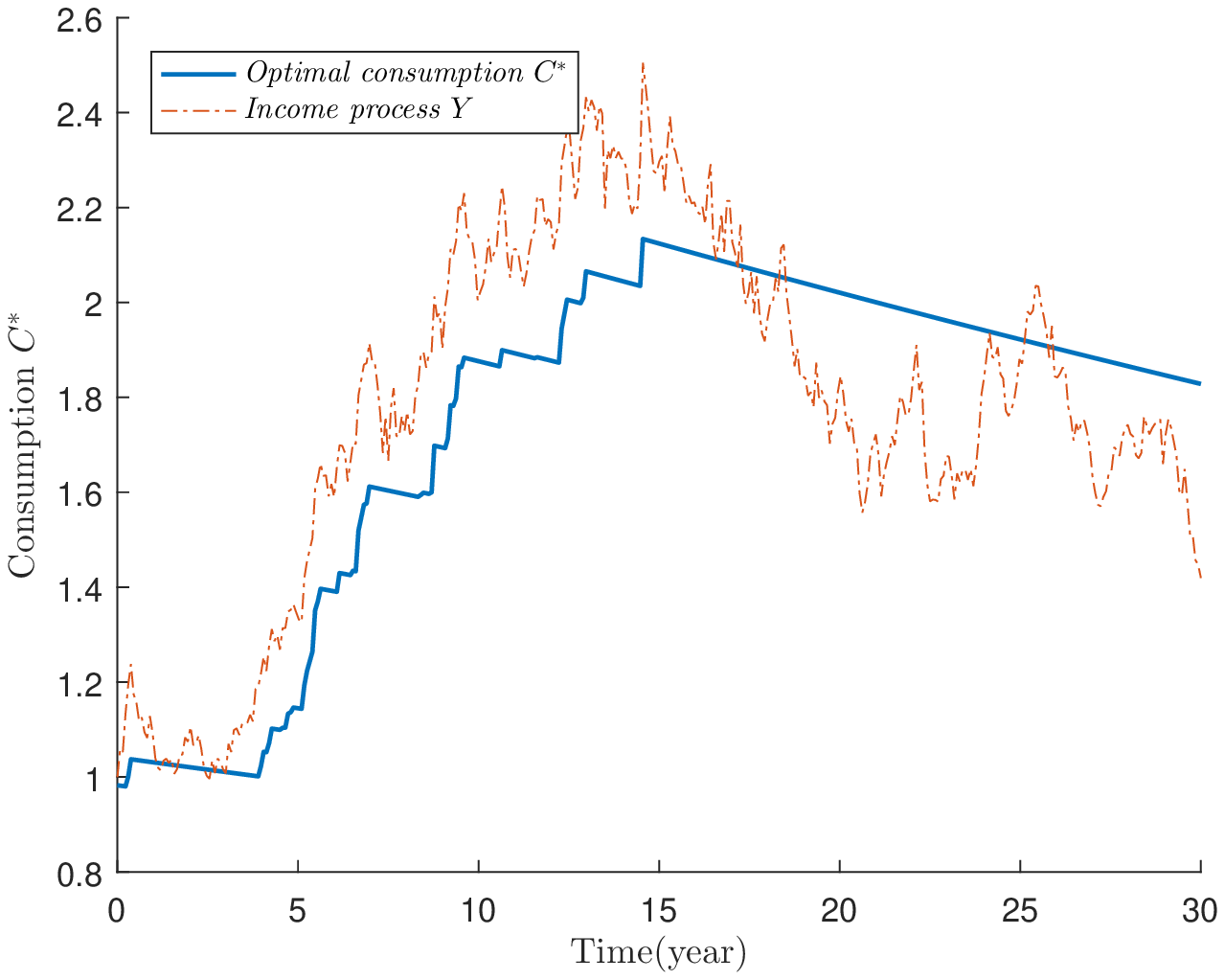}}
	\caption{\label{fig:3} Simulation of optimal consumption $C^*$. The parameter values are as follows: $\rho=0.07,\;r=0.04,\;\mu=0.02\;\sigma=0.1,\;\gamma=3\;w=-5$ and $T=30$.}
\end{figure}

Figure \ref{fig:3} which is representative of the case where $\rho>r${,} show {ing} a simulation path of optimal consumption allocation. In the case of $\rho>r$, however, the optimal consumption $C^*$ gradually decreases even when the regulated process $\lambda^*/Y^\gamma$ does not hit the free boundary.
\begin{figure}[h]
	\centering
	\subfigure[Scenario 1]{\label{fig01a}\includegraphics[scale=0.45]{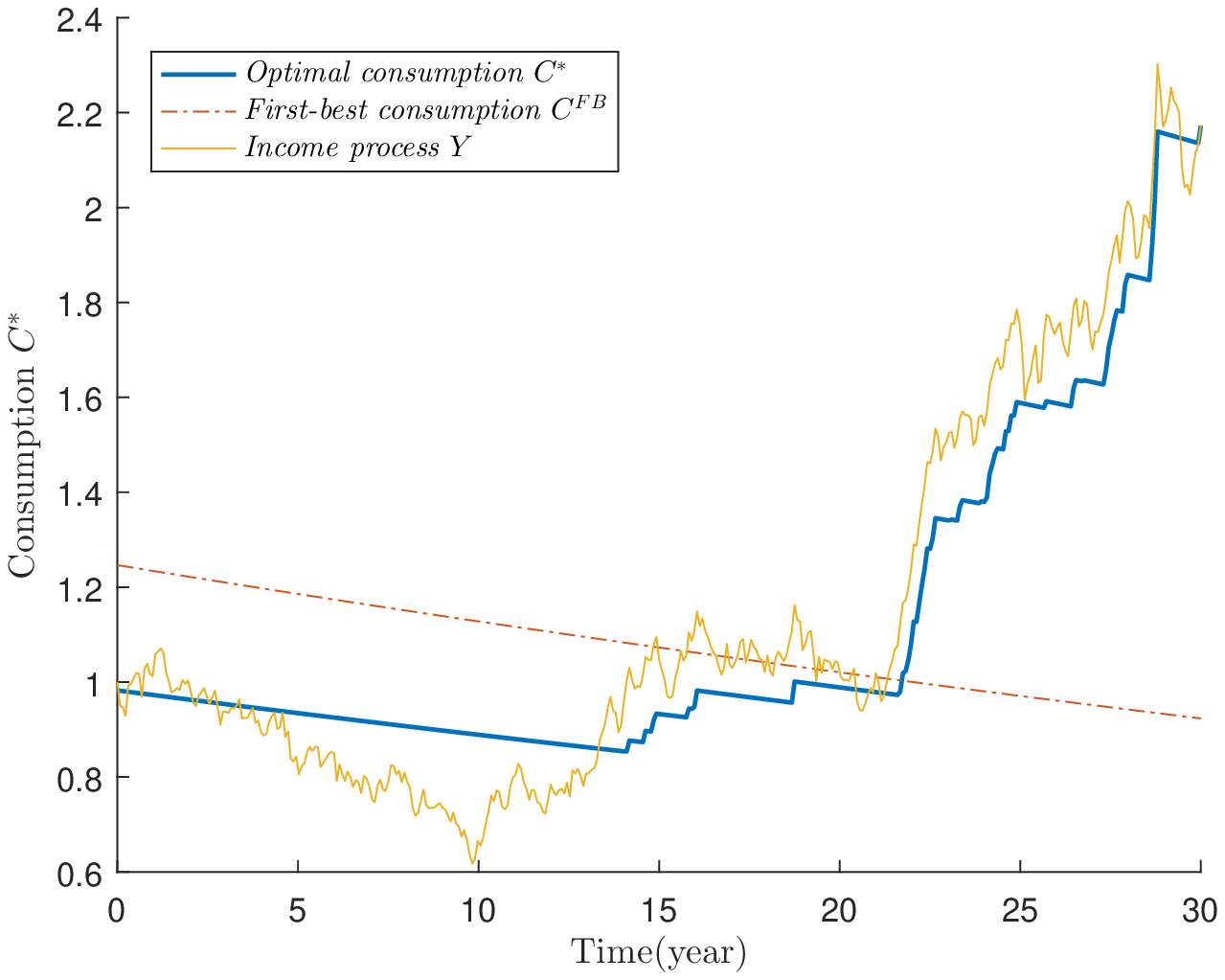}}
	\subfigure[Scenario 2]{\label{fig01a}\includegraphics[scale=0.45]{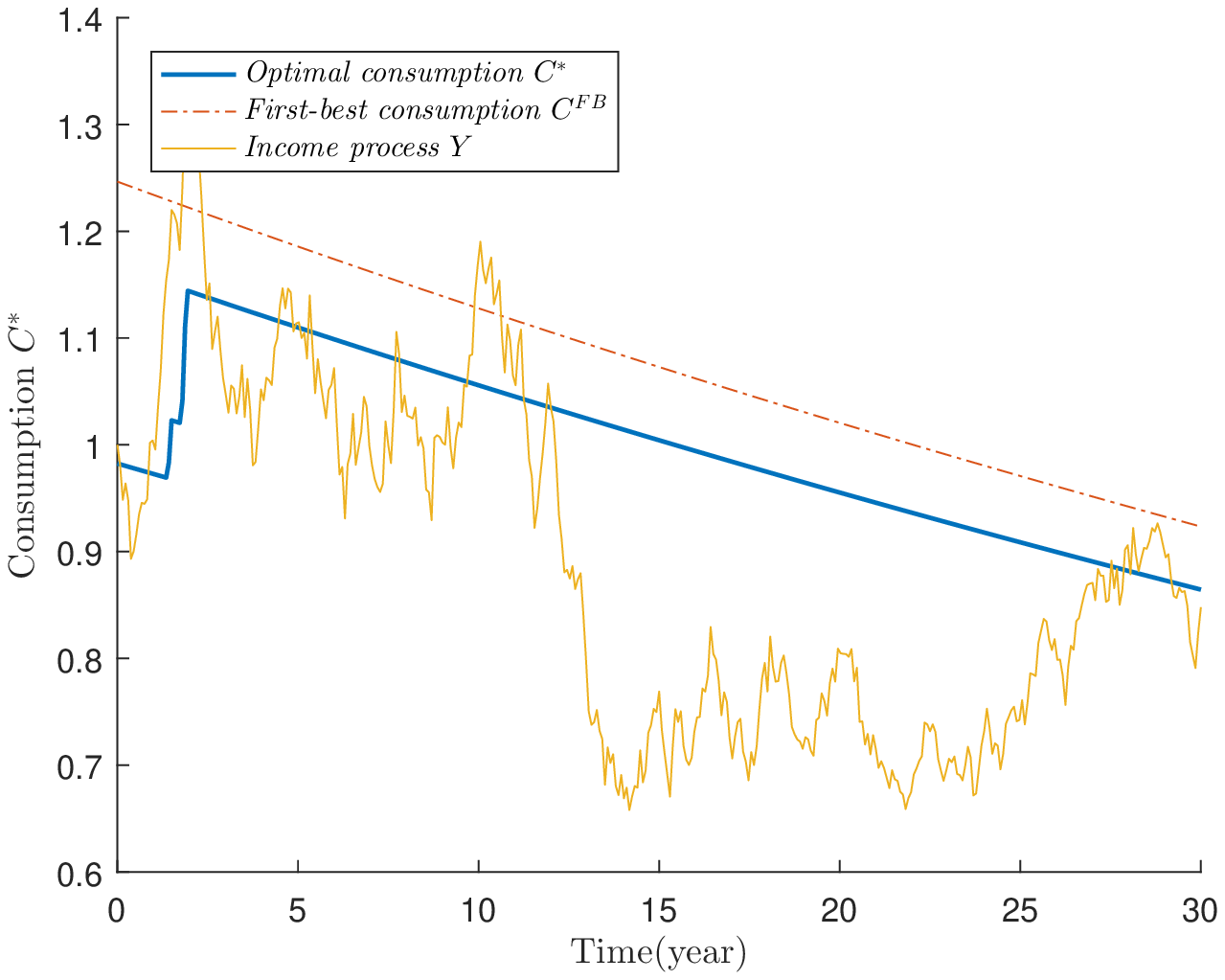}}
	\caption{\label{fig:4} Simulation of optimal consumption $C^*$. The parameter values are as follows: $\rho=0.07,\;r=0.04,\;\mu=0.02\;\sigma=0.1,\;\gamma=3\;w=-5$ and $T=30$.}
\end{figure}

Figure \ref{fig:4} shows the comparison results of the optimal consumption $C^*$ and the first-best consumption $C^{FB}$ in two different scenarios. Two scenarios correspond to two different sample paths of the income process generated with the same parameter values. In the first scenario (scenario 1), the income process steadily increases whereas the income process tends to decrease in the second scenario (scenario 2). 
Since the first-best consumption defined in \eqref{eq:FB_C} is a deterministic function, $C^{FB}$ is the same in both scenarios. In contrast to the first-best allocation, the optimal consumption $C^*$ in limited commitments depends on  {the entire} history of income process $Y_t$. 
In other words, the optimal consumption $C^*$ also increases as income steadily increases in the scenario 1. That is, although $C^* <C^{FB}$ at the beginning, as time passes, the optimal consumption $C^*$ exceeds the first-best allocation $C^{FB}${,} as shown in Figure \ref{fig:4} (a). In scenario 2, however, since the income process $Y$ gradually decreases, the optimal consumption $C^*$ does not exceed first-best allocation $C^{FB}$ until maturity $T$. 
\section{Concluding Remarks}\label{sec:5}

In this paper we study the optimal contracting problem with limited commitment between a risk-neutral principal and a risk-averse agent in the finite horizon. We establish the duality relationship and transform the dual problem into an infinite series of optimal stopping problems, which essentially becomes a single optimal stopping problem. This is similar in its formal structure to an irreversible incremental investment problem studied in \citet{Pindyck88}  {and} \citet{DixitPindyck}.  {B}ased on partial differential equation theory, we characterize the variational inequality arising from the optimal stopping problem. We also derive an integral equation representation of optimal strategy and provide numerical results  {for} the optimal strategy. 
\section*{Acknowledgments.}

Junkee Jeon gratefully acknowledges the support of the National Research Foundation of Korea (NRF) grant funded by the Korea government (Grant No. NRF-2017R1C1B1001811). Hyeng Kuen Koo gratefully acknowledges the support of the National Research Foundation of Korea (NRF) grant funded by the Korea government (MSIP) (Grant No. NRF-2016R1A2B4008240). Kyunghyun Park is supported by NRF Global Ph.D Fellowship (2016H1A2A1908911).





\end{document}